\documentclass[aps,pre,showpacs,floatfix,twocolumn,amsmath,amssymb,dvips]{revtex4-1}%
\usepackage{amsmath}
\usepackage{epsfig}
\usepackage{graphicx}
\usepackage{dcolumn}
\usepackage{bm}
\usepackage{amsfonts}
\usepackage{amssymb}%
\usepackage[colorlinks=true,dvipdfm]{hyperref}

\begin{document}
\title{Renormalization-group theory for temperature-driven first-order phase transitions in scalar models}
\author{Ning Liang and Fan Zhong}
\thanks{Corresponding author. E-mail: stszf@mail.sysu.edu.cn}
\affiliation{State Key Laboratory of Optoelectronic Materials and Technologies, School of
Physics and Engineering, Sun Yat-sen University, Guangzhou 510275, People's
Republic of China}

\date{\today }

\begin{abstract}
We study the scaling and universal behavior of temperature-driven first-order phase transitions in scalar models. These transitions are found to exhibit rich phenomena, though they are controlled by a single complex-conjugate pair of the imaginary fixed points of a $\phi^3$ theory. Scaling theories and renormalization-group theories are developed to account for the phenomena. Several universality classes with their own hysteresis exponents are found including a field-like thermal class, a partly thermal class, and a purely thermal class, designated respectively as Thermal Class I, II, and III. The first two classes arise from the opposite limits of the scaling forms proposed and may cross over to each other depending on the temperature sweep rate. They are both described by a massless model and a purely massive model, both of which are equivalent and are derived from the $\phi^3$ theory via symmetry. Thermal Class III characterizes the cooling transitions in the absence of applied external fields and is described by purely thermal models, which includes cases in which the order parameters possess different symmetries and thus exhibiting different universality classes. For the purely thermal models whose free energies contain odd-symmetry terms, Thermal Class III emerges only in mean-field level and is identical with Thermal Class II. Fluctuations change the model into the other two models. Using the extant three- and two-loop results for the static and dynamic exponents for the Yang-Lee edge singularity, respectively, which falls into the same universality class to the $\phi^3$ theory, we estimate the thermal hysteresis exponents of the various classes to the same precisions. Comparisons with numerical results and experiments are briefly discussed.
\end{abstract}

\maketitle

\section{\label{intro}Introduction}

Many first-order phase transitions (FOPTs) are driven to occur by varying the
temperature $T$ and thermal hysteresis often ensues. The energy dissipations
in the processes studied by internal frictions~\cite{Lin} and by thermal
analysis~\cite{Zhang,liu,fung,Kuang} were found to follow a power law with respect to the
sweep rate $R$ of the temperature. On the basis of the time-dependent
Ginzburg-Landau theory of a $(\phi^{2})^{3}$ model with an $O(N)$ symmetry in the vector order parameter $\phi$ space in the limit of large components $N$, it was
found that the areas $A$ of the thermal hysteresis loops under a sinusoidally
varying temperature depend on the amplitude $r_{a}$ of the variation as
$A=\oint MdT\propto r_{a}^{\Upsilon}$ with a hysteresis exponent $\Upsilon%
=1.0\pm0.03$~\cite{Rao}. By using
a linearly rather than sinusoidally varying temperature with a constant rate $R$, which is
experimentally more amenable, familiar spindle-shaped thermal hysteresis loops
of both the large-$N$ $(\phi^{2})^{3}$ and mean-field models were
reached~\cite{zhong}. In an entropy $S$ versus
$T$ frame, the areas of the loops, which are proportional to the energy dissipations in the cycles, were found to be
\begin{equation}
A_{2}=A_{0}+aR^{\Upsilon}, \label{a}%
\end{equation}
with $\Upsilon=2/3$ independent of the model parameters, where $A_{0}$ and $a$
are constants~\cite{zhong}. This dynamic hysteresis scaling including the exponent has been confirmed
experimentally in a nematic--smectic-$A$ phase transition of a binary mixture
under a linearly varying temperature~\cite{Yildiz}. Recently, it was shown by
a renormalization-group (RG) theory that the FOPTs driven by an applied external field in a scalar $\phi^{4}$
model \emph{below} its critical temperature are controlled by the instability fixed
point of a derived $\phi^{3}$ model~\cite{zhongchen,Zhong2}. Although it is
imaginary in value, the fixed point is physical counter-intuitively in order
for the $\phi^{3}$ theory to be mathematically convergent~\cite{zhonge12}. Numerical evidence from the Potts model for the fixed point has also been found~\cite{Fan}. A functional RG theory has extended the theory to contain all odd-order terms and given rise to consistent exponents~\cite{li12}. These then place
the scaling behavior of the FOPTs in the same universality class to the
Yang-Lee edge singularity~\cite{Fisher}, the singularity of the distribution of
the Yang-Lee zeros \emph{above} the critical point.

In the $\phi^3$ theory~\cite{zhongchen,Zhong2}, one first notices that, for an FOPT, there exists a sharply defined curve of spinodals at which the associated susceptibility diverges on the mean-field level that becomes exact for long ranged interactions. Upon an expansion near them, these spinodals manifest themselves as the ``critical'' point of a $\phi^3$ theory, because the mean-field free energy now contains a $\phi^3$ term as the leading coupling term even if the symmetry of the original theory excludes it. As the $\phi^3$ theory is unstable at its critical point, we dub it instability point and the critical exponents characterizing it instability exponents. Although one has a $\phi^3$ theory for the unstable phenomena instead of the usual $\phi^4$ one for critical phenomena, one can easily convince oneself, as will be seen in the sections following for the temperature-driving transitions, that the whole theory for the latter applies to the former as well~\cite{Zhong2}. This includes the definitions (or the meaning) and computations of the mean-field instability exponents, the fluctuations that lead to a divergent correlation length, the Ginzburg criterion for the relative importance of the fluctuations, the shift of the instability point due to the fluctuations, and the RG theory for the non-Gaussian fixed point and its associated non-mean-field instability exponents, which combine to characterize the hysteresis near the instability point, as well as the irrelevance of the neglected higher order terms~\cite{Zhong2}.

However, there exist several possible questions to the $\phi^3$ theory. The first one concerns the physical meaning of the instability points and the existence of the spinodals. For a system with short-ranged interactions, it is generally believed that there exist no sharply defined spinodals in contrary to the mean-field case with long-ranged interactions~\cite{Gunton,Binder1,Binder2}. This can be readily understood from the fact that thermal fluctuations simply smear any possible spinodal out into a region when the nucleation barrier is on the order of the thermal energy. Emphasizing the effect of the fluctuations nevertheless, the existence of a crossover between nucleation and spinodal decomposition is out of question as the former needs activation whereas the latter does not at least away from the crossover region and at the early time of the transition. It is then more than natural to postulate the existence of hidden spinodals which, albeit smeared out again by the thermal fluctuations, control the crossover. These hidden spinodals, which are shifted by the fluctuations away from their mean-field values, are nothing but the shifted instability points. Their magnitudes are non-universal and depends on the coarse grained scale, in agreement with numerical results~\cite{Binder78,Kawasaki,Kaski,Gunton}. Although studies were performed that vary the range of interactions to approach the mean-field spinodals~\cite{Herrmann,Gulbahce}, from the analogous $\phi^4$ theory for criticality, it is clear that the mean-field theory is only the correct starting point for the RG analysis of fluctuations in systems with short-ranged interactions. There is no apparent contradiction at all in describing the critical (unstable) behavior of a short-ranged-interaction system around a particular point existing only within the long-ranged-interaction mean-field models.

Another question pertains to the only prominent difference between the $\phi^3$ and the $\phi^4$ theories, viz., the fixed point of the former is imaginary in value. However, it has been found that for a physical coupling of a purely real initial value, the RG flow has to diverge at a finite scale in order to achieve an imaginary part~\cite{Zhong2}. Upon combining a momentum-shell integration RG analysis and a nucleation theory \cite{Langer67} near the spinodal point \cite{Klein83,Unger84}, it has been shown that~\cite{zhonge12}, exactly at the finite scale, vanishes the free-energy cost for nucleation out of the metastable state in which the system lies, together with the potential well of the metastable state itself. This places the system exactly at a true instability point and thus exhibiting the divergence at the scale. The integration for the partition function then diverges and has to be analytically continued to the complex plane in order to be physically meaningful. As a consequence, the system enters the imaginary domain and can thus reach the imaginary fixed point. Therefore, counter to the intuition that only real values are physical, for the $\phi^3$ theory, imaginary values are physical instead~\cite{zhonge12}.

Yet another possible question is the relation to the nucleation and growth~\cite{Gunton,Binder1,Oxtoby}. As the theory is an expansion around the instability points, it can apparently describe the behavior of spinodal decomposition~\cite{Gunton} at least at its early time. We note that, as the other dynamical mode of FOPTs, nucleation is the classical theory for FOPTs and has been argued to result in, for small $R$, a hysteresis which vanishes in a purely logarithmic form~\cite{Thomas}. Although it has been shown unambiguously that this non-perturbative nucleation effect can only play a role at extremely low rates which are not accessible experimentally~\cite{Sides98,Sides99}, no evidence of an overall power-law relationship has been found for the magnetic hysteresis in a sinusoidally oscillating field in two dimensions either~\cite{Sides98,Sides99}. A direct consequence of these nucleation theory is that the hysteresis loop ought to shrink and vanish in the limit of small rates. However, numerical simulations of the Ising model found existence of a dynamic phase transition at finite amplitudes of the field, the transition below which the switch between the two phases of opposite magnetizations cannot take place~\cite{Acharyya,Sides98,Sides99}. It was even claimed that a finite field is needed to flip the magnetization even in the static limit of the field~\cite{Acharyya}. These indicate that nucleation alone cannot be all the story and the loci at which the dynamic transition occurs may well be the dynamic instability points even in the regime in which nucleation is thought to be dominant. The combination of the RG and the nucleation theory mentioned above shows that~\cite{zhonge12} the characteristic length scale, at which the RG flows diverge and become imaginary, divides the fluctuating field of a metastable system into two sets. The short modes feel a finite potential well and are responsible for nucleation, whereas the long ones have none and are controlled by the imaginary fixed point. Although the short modes are renormalized away when coarse-grained and are thus irrelevant in the sense of the RG theory, they may well compete with the long ones in a real system itself in which no renormalization is performed and mask the scaling behavior. Physically in a real system, when the time scale of nucleating out of the metastable state is shorter than that of the driving from, say, the equilibrium transition point to near the instability point, nucleation will probably occur. But even near the instability point where nucleation barriers still exist owing to the $\phi^4$ interaction that is irrelevant and neglected in the $\phi^3$ theory, nucleation may still play a role. Accordingly, how to disentangle the nucleation and the scaling from the $\phi^{3}$ theory is yet to be resolved. It is therefore helpful to turn to more examples.

Here, we shall develop RG theories for the dynamics of temperature-driven
FOPTs in scalar models. We find that although they are driven by varying the temperature instead of an external field, these FOPTs are
again controlled by the same instability fixed point of the $\phi^{3}$ theory
for field-driven transitions. Distinct from the latter case, however, there
are now complicate behavior. The temperature-driven FOPTs exhibit three universality classes with three sets of thermal hysteresis exponents. We call them Thermal Class I, II, and III, corresponding respectively to a field-like thermal class, a partly thermal class, and a purely thermal class. There is also a crossover between the first two classes. The last class can have further variants depending on the symmetry of the model. All these rich behaviors are clearly demonstrated from reduced models and well accounted for by the RG theories.

In the following, we shall first study mean-field models for the thermal
transitions in Sec.~\ref{model} and identify the three different universality classes in Sec.~\ref{mft}. Various reduced models for the different classes are also derived in the section.
A scaling theory is then developed in Sec.~\ref{scaling} to account for the
peculiar scaling behavior. The RG theories for them follows in Sec.~\ref{rgt}. A
summary is given in Sec.~\ref{summary}.

\section{\label{model}Model}

As a simple model that involves an FOPT between an ordered phase at low
temperatures and a disordered phase at high temperatures, we consider the
following free-energy functional \cite{landau,gennes} in a $d$-dimensional
space
\begin{equation}
F[\phi]=\int{d\mathrm{\mathbf{r}}\left\{  {\frac{1}{2}r\phi^{2}+\frac{1}%
{2}[\nabla\phi]^{2}+\frac{1}{3!}w\phi^{3}+\frac{1}{4!}g\phi^{4}-H\phi
}\right\}  }, \label{f}%
\end{equation}
where $r$ is a reduced temperature proportional to $T$, $H$ an external field
conjugated to the scalar order parameter field $\phi$, while $w$ and $g $ are
coupling constants. For stability, $g>0$. Yet, $w$ can be either positive or negative. For simplicity, we shall choose $w<0$ throughout, which results in an ordered phase with a positive equilibrium order parameter. Equation~(\ref{f}) contains a cubic term and thus $\phi$ has no inversion symmetry. If this symmetry is respected, on the other
hand, a $\phi^{6}$ term has to be included in order to have a thermal FOPT
\cite{Devonshire}. However, it will be seen later on that the transition
is still governed by the same theory to be developed below except the cooling transition in $H=0$, which belongs to the purely thermal class. We thus focus here on the model defined by Eq.~(\ref{f}).

To study the transition driven to metastable states as the temperature is
varied, dynamics has to be taken into account. We employ a purely relaxation
dynamics of a non-conserved order parameter, i.e., Model A \cite{Hohenberg},
described by the Langevin equation
\begin{equation}
\frac{\partial\phi}{\partial t}=-\lambda\frac{\delta F[\phi]}{\delta\phi}
+\zeta, \label{lang}%
\end{equation}
where $\lambda$ is a kinetic coefficient and the Gaussian white noise $\zeta$
satisfies
\begin{align}
\langle\zeta(\mathbf{r},t)\rangle &  =0,\nonumber\\
\langle\zeta(\mathbf{r},t)\zeta(\mathbf{r^{\prime}},t^{\prime})\rangle &  =
2\lambda T\delta(\mathbf{r}-\mathbf{r^{\prime}})\delta(t-t^{\prime}),
\label{noise}%
\end{align}
and mimics an effect of other degrees of freedom, where the angle brackets
denote averages over the noise.

\section{\label{mft} MEAN-FIELD THEORY: Reduced models and thermal classes}

\subsection{Theory and reduced models}

\begin{figure}
\includegraphics[width=8.5cm]{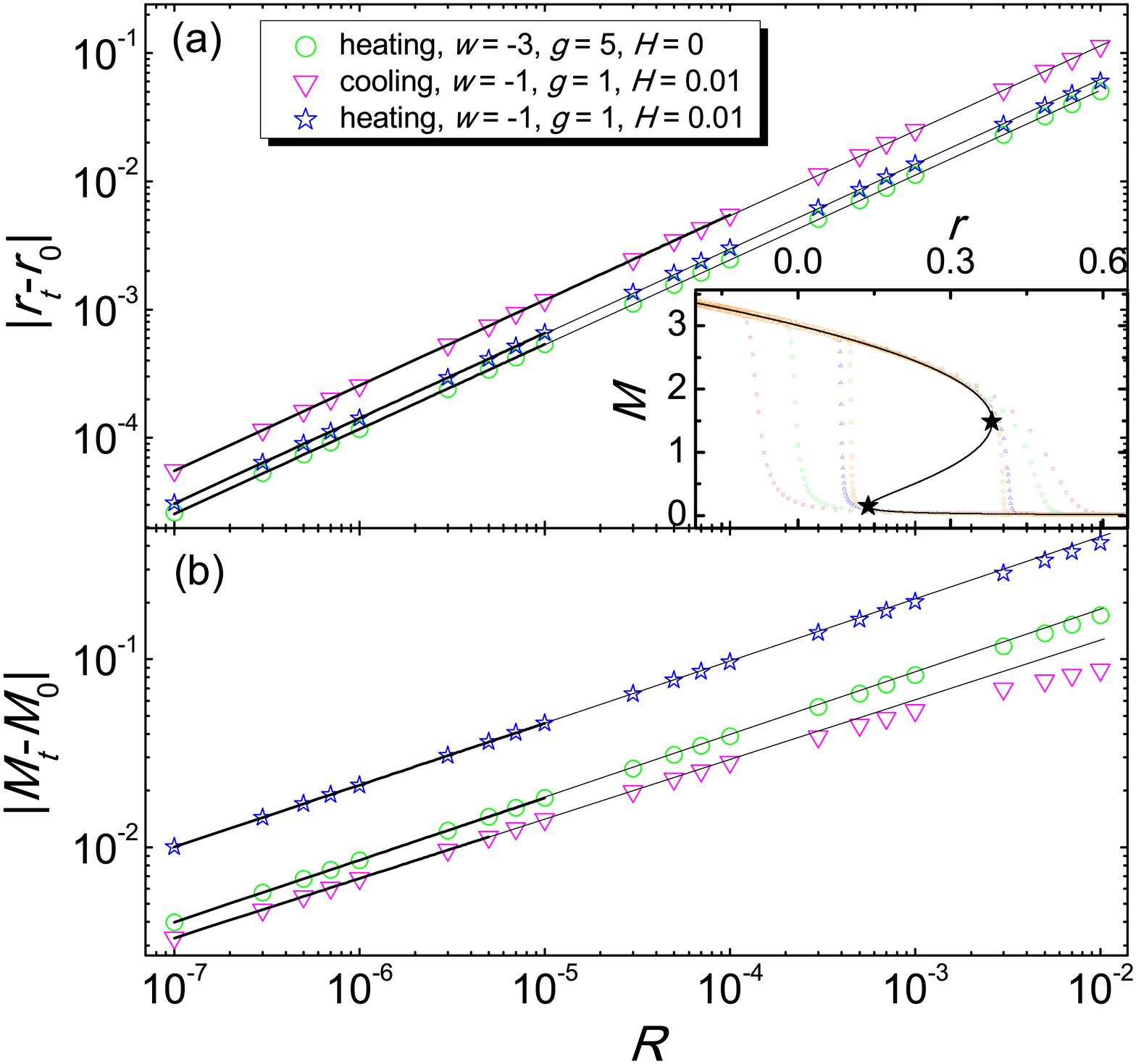}\caption{(Color online) Dependence of (a) the transition temperature $r_{t}$ and (b) the transition order parameter $M_t$ on the temperature sweep rate
$R$. The thick lines are fits to the data covered and the thin lines are their direct extensions. The fitted lines have slopes $0.6649(2)$, $0.6645(8)$, $0.664(2)$ and intercepts $0.1379148(3)$, $0.3816972(3)$, $0.674999(4)$ in (a) and $0.3286(3)$, $0.3303(2)$, $0.3167(8)$ and $1.48627(2)$, $0.899960(6)$, $0.14917(2)$ in (b) from up to down. The corresponding theoretical slopes are $2/3$ in (a) and $1/3$ in (b), and the corresponding theoretical intercepts are $0.1379139$, $0.3816968$. $0.675$ in (a) and $1.48642$, $0.9$, $0.149017$ in (b). Inset: Hysteresis loops for several $R$ of the parameter set with a finite field. The two black stars mark the spinodal/instability points. The curve connecting the two spinodal points is Eq.~(\protect\ref{df}).}
\label{rMs}
\end{figure}

In the mean-field approximation, all fluctuations are ignored. As
a result, $\langle\phi^k\rangle=M^k$ and Eq.~(\ref{f}) reduces to
\begin{equation}
\frac{\partial M}{\partial t}=-\lambda\left(  rM-{\nabla}^{2}{M}+\frac{1}%
{2!}wM^{2}+\frac{1}{3!}gM^{3}-{H}\right), \label{Me}%
\end{equation}
where we have kept the gradient term for later uses though $M$ is spatially uniform. In
equilibrium and $H=0$, Eq.~(\ref{Me}) describes an FOPT at an equilibrium
transition temperature $r_{e}=w^{2}/3g$, at which $M$ jumps from a disordered
phase with $M=0$ to an ordered phase with $M=M_{e}=(-3w+\sqrt{9w^{2}-24rg})/2g$. The stability limits or the spinodal points of the two phases lie at $(r_{s0}
^{-},M_{s0}^{-})=(0,0)$ and $(r_{s0}^{+},M_{s0}^{+})=(3w^{2}/8g,-3w/2g)$,
respectively, which are solutions of
\begin{subequations}
\label{spinodal}%
\begin{align}
r_{s0}M_{s0}+\frac{1}{2}wM_{s0}^{2}+\frac{1}{3!}gM_{s0}^{3}-H  &  =0,\label{df}\\
r_{s0}+wM_{s0}+\frac{1}{2}gM_{s0}^{2}  &  =0, \label{ddf}%
\end{align}
with $H=0$. We shall refer to the two spinodal points as cooling and heating spinodal/instability points as they are relevant to cooling and heating transitions, respectively. For $H\neq0$, below the critical point $H_{c}=-w^{3}/6g^2$,
$r_{c}=w^{2}/2g$, and $M_{c}=-w/g$ given by Eq.~(\ref{spinodal}) and the derivative
of Eq.~(\ref{ddf}) with $M_{s0}$ (with all variables replaced by the critical ones), there are also FOPTs from a phase with a
small $M$ to that with a large $M$. The spinodal points in this case are also determined
by Eq.~(\ref{spinodal})~\cite{zhong}.

As there are no fluctuations, the dynamic FOPTs described by Eq.~(\ref{Me})
can only take place beyond $(r_{s0},M_{s0})$ as shown in the inset of
Fig.~\ref{rMs}, because at the spinodal the barrier between the two phases vanishes and the system becomes unstable. Accordingly, we set
\end{subequations}
\begin{equation}
M(t)=M_{s0}+m(t), \label{mshift}%
\end{equation}
and $m(t)$ is then described by the dynamics
\begin{equation}
\frac{\partial m}{\partial t}=-\lambda\left(  \tau m-{\nabla}^{2}{m+}\frac
{1}{2!}vm^{2}-K\right)  , \label{mnew}%
\end{equation}
with
\begin{subequations}
\label{taug}%
\begin{align}
\tau &  =r+wM_{s0}+\frac{1}{2}gM_{s0}^{2}=r-r_{s0},\label{tau}\\
K  &  =H-rM_{s0}-\frac{1}{2}wM_{s0}^{2}-\frac{1}{3!}gM_{s0}^{3}=-M_{s0}\tau,\label{tau2}\\
v  &  =w+gM_{s0}, \label{g}%
\end{align}
\end{subequations}
near $(r_{s0},M_{s0})$ from Eqs.~(\ref{Me}) and (\ref{spinodal}). Note that although
Eq.~(\ref{taug}) is exact, in Eq.~(\ref{mnew}), we have kept only the leading
$m^{2}$ term and neglected the higher-order $m^{3}$ term for $(r,M)$
sufficiently near $(r_{s0},M_{s0})$ and hence small $\tau$ and $m$. As a consequence,
the free-energy functional responsible for Eq.~(\ref{mnew}) is a
$\phi^{3}$ theory in an effective field $K$. Therefore, the dynamics of the
temperature-driven FOPTs near their instability points $\tau=0$ and $K=0$ is governed by the $\phi^{3}$ theory similar to the field-driven FOPTs.

The $\phi^{3}$ theory bears a particular symmetry that simplifies the theory.
The model as defined in Eq.~(\ref{mnew}) appears to have two parameters $\tau$
and $K$ for controlling its distance to the instability point. However, it is
well known that there is only one independent parameter for a $\phi^{3}$
theory~\cite{Fisher}. Indeed, since a shift of the order parameter by a constant amount $c$,
i.e.,
\begin{equation}
m=\varphi-{c}, \label{shift}%
\end{equation}
only changes
\begin{subequations}
\label{tkprimed}%
\begin{align}
m  &  \rightarrow\varphi,\nonumber\\
\tau &  \rightarrow\tau^{\prime}=\tau-{v}c,\label{tauprimed}\\
K  &  \rightarrow K^{\prime}=K+\tau c-\frac{1}{2}{v}c^{2}, \label{kprimed}%
\end{align}
\end{subequations}
and thus keeping the structure of Eq.~(\ref{mnew}), a particular choice of $c$
can turn the mass term $\tau$ into the effective-field term $K$ and vice
versa~\cite{Breuer}. Accordingly, we have two equivalent reduced $\phi^{3}$ theories
to describe the thermal transitions.

The first one is a massless model of
\begin{equation}
\frac{\partial\varphi}{\partial t}=-\lambda\left(  -{\nabla}^{2}\varphi
{+}\frac{1}{2!}{v}\varphi^{2}-\overline{K}\right)  \label{Langevin 2}%
\end{equation}
with a shifted effective field (denoted by an overline)
\begin{equation}
\overline{K}=K+\tau^{2}/2{v}=-M_{s0}\tau+\tau^{2}/2{v} \label{Kbar}%
\end{equation}
for $c=\tau/{v}$.

The second is a purely massive model of
\begin{equation}
\frac{\partial\varphi}{\partial t}=-\lambda\left(  \overline{\tau}\varphi-{\nabla
}^{2}\varphi{+}\frac{1}{2!}{v}\varphi^{2}\right)  \label{Langevin 3}%
\end{equation}
with a shifted mass satisfying
\begin{equation}
\overline{\tau}^{2}=\tau^{2}+2vK=\tau(\tau-2vM_{s0})\label{taubar}%
\end{equation}
for $c=\tau/v+\sqrt{\tau^{2}+2vK}/v$. However, this massive model is real only near the cooling spinodals $(r_{s0}^{-},M_{s0}^{-})$ in $H=0$. This can be seen as follows. From Eq.~(\ref{g}), we see that for $M_{s0}=M_c$, $v(M_c)=0$. Because the heating and the cooling spinodals merge at $M_c$ and are all positive for $w<0$ chosen, for the heating spinodals, $M_{s0}>M_c$ and so $v>0$ and hence $2vM_{s0}>0$; while for the cooling spinodals, $M_{s0}<M_c$ and thus $v<0$ and hence $2vM_{s0}<0$. The heating spinodals are the stability limits of the ordered phases and are relevant for the heating transitions. This means that $r>r_{s0}$ and $\tau>0$ near the heating spinodals. A positive $2vM_{s0}$ then implies that $\overline{\tau}^2$ must be negative for sufficiently small heating rates and hence sufficiently small $\tau$ from Eq.~(\ref{taubar}). Indeed, in $H=0$, for instance, a direct computation gives $2vM_{s0}^+=4r_{s0}^+>0$ and $\overline{\tau}^{2}=(r-r_{s0}^+)(r-5r_{s0}^+)$, which is negative for the physically relevant range of $r_{s0}^+<r<5r_{s0}^+$. On the other hand, the cooling spinodals are relevant for the cooling transitions and thus $r<r_{s0}$ and $\tau<0$ near them. As a result, a negative $2vM_{s0}$ implies that $\overline{\tau}^2$ must be again negative for sufficiently small cooling rates.
Yet, there is a special case in which $\overline{\tau}$ is real. This is the case for the cooling spinodals in $H=0$, in which $M_{s0}^-=0$ and thus $\overline{\tau}$ is simply $\tau$. Consequently, Eq.~(\ref{Langevin 3}) is just Eq.~(\ref{Me}) with the cubic term omitted. Also, in this case, $c=0$, which is the reason why we have kept only the plus sign for it.
We see therefore that most of the temperature-driven transitions in the mean-field theory cannot be cast into a purely massive theory with real parameters. Nevertheless, we still study it as multiplying by the imaginary unit $i$ changes formally the theory to that of the Yang-Lee edge singularity, which falls into the same universality class as mentioned.

The foregoing discussions indicate that besides the two equivalent reduced models~(\ref{Langevin 2}) and (\ref{Langevin 3}), there exists another purely massive model
\begin{equation}
\frac{\partial\varphi}{\partial t}=-\lambda\left(  \tau\varphi-{\nabla
}^{2}\varphi{+}\frac{1}{2!}{v}\varphi^{2}\right),  \label{ptm}%
\end{equation}
which we call purely thermal model. The only difference between the two purely massive models is that the mass term is $\tau$ instead of $\overline{\tau}$ for the purely thermal model because the order parameter at the instability point vanishes. As mentioned, the model~(\ref{ptm}) describes the cooling transition in $H=0$ in the mean-field approximation. However, we shall see below that this is true only in mean field. When fluctuations are taken into account, an effective field $K$ is generated. As a consequence, we have to resort to the other two models.

The purely thermal model~(\ref{ptm}) can be generalized to a general form,
\begin{equation}
\frac{\partial\varphi}{\partial t}=-\lambda\left(  \tau\varphi-{\nabla
}^{2}\varphi{+}\frac{1}{\sigma!}{v}\varphi^{\sigma}\right),  \label{Langevin 4}%
\end{equation}
with an integer $\sigma$. For $\sigma=2$, it recovers the purely thermal model above. For $\sigma=3$, on the other hand, it describes the cooling transition of the $\phi^6$ model at $H=0$. It is also equivalent to the usual $\phi^4$ model for continuous phase transitions~\cite{zhonge}. We thus expect the two models with a different $\sigma$ to fall into different universality classes. It is straightforwardly to extend the analysis to bigger values of $\sigma$, but we shall not consider them further.

If the Gaussian noise~(\ref{noise}) is reintroduced back, all the four reduced models can describe the
fluctuations near the instability points. The massless model
was studied in~\cite{zhongchen,Zhong2} and the equilibrium properties
of the purely thermal model was investigated in~\cite{zhonge12}. For temperature-driven FOPTs, however, the controlling parameter in
the first two models is a nonlinear function of $\tau$, the
reduced temperature to the instability point. We
shall see that it is this nonlinearity that leads to the complication of the
dynamic scaling for such transitions. The purely thermal model with an even $\sigma$ in the presence of fluctuations also returns to the first two models. Only the generalized purely thermal model with an odd $\sigma$ has simple behavior due to its free of the nonlinearity. In fact, it gives rise to the purely thermal class of Thermal Class III, which also results from the purely thermal model with an even $\sigma$ in mean field, while the first two models result in the other two classes and their crossover in some special limits.
In the following, when we deal with the purely massive models, we mainly consider Eq.~(\ref{Langevin 3}), as the results for the purely thermal model Eq.~(\ref{ptm}) can be obtained by direct replacements.

We emphasize that the reduced models can describe the thermal transitions effectively because they only contain one controlling parameter. The general model~(\ref{mnew}) contains explicitly one redundant parameter and care has to be executed when it is used.

\subsection{Thermal Classes}
We identify the thermal classes in this subsection from numerical and some possible analytical results.

\subsubsection{Field-like Thermal Class I}

The usual dynamic scaling described by Eq.~(\ref{a}) with $\Upsilon=2/3$ for the field-like Thermal Class I can be readily found from finite-time scaling (FTS) using a linearly varying
temperature~\cite{gong,zhongfts} by numerically solving Eq.~(\ref{Me}) with $r=r_{s0}+Rt$.
We have purposely chosen the time origin at the instability point for
simplicity. In fact, once it is sufficiently far away from $r_{s0}$, the
initial $r$ and hence the time origin have no effects on the scaling. The
reduced transition temperature $r_{t}$ characterized by the $r$ at $M=M_{s0}$ follows
\begin{equation}
r_{t}=r_{0}+a_{1}R^{\Upsilon}, \label{rscale}%
\end{equation}
with the thermal hysteresis exponent $\Upsilon=2/3$ for sufficiently small
$R$ similar to Eq.~(\ref{a}), where $r_{0}$ and $a_{1}$ are constants. It is
found that the smaller the $R$ values are, the closer $\Upsilon$ to $2/3$ as
shown in Fig.~\ref{rMs}(a). Similarly, the transition order parameter $M_{t}$ at
$r=r_{s0}$ obeys
\begin{equation}
M_{t}=M_{0}+a_{2}R^{\Upsilon_{m}}, \label{mscale}%
\end{equation}
with the order-parameter hysteresis exponent $\Upsilon_{m}=1/3$ for sufficiently
small $R$, as shown in Fig.~\ref{rMs}(b). One sees there that $r_{0}$ and $M_{0}$ are $r_{s0}$ and $M_{s0}$, respectively.

\begin{figure}
\includegraphics[width=8.5cm]{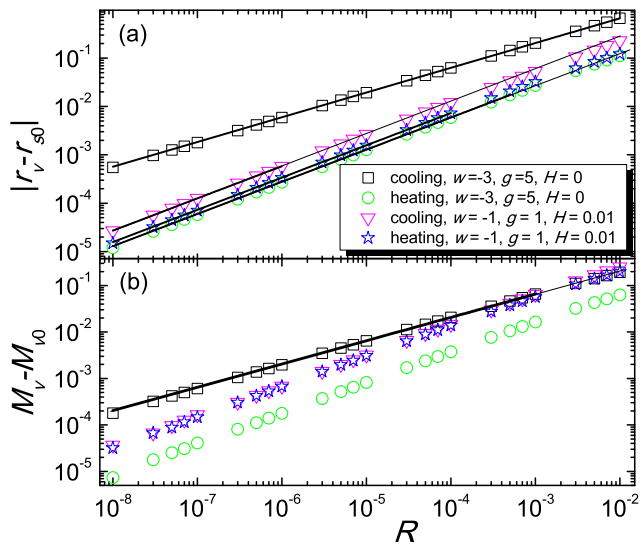}\caption{(Color online) Dependence of (a) the temperature $r_{v}$ and (b) the order parameter $M_v$ at the nadir of the valley of $dM/dr$ on the temperature sweep rate $R$. The thick lines are power-law fits to the data covered and the thin lines are their direct extensions. The fitted lines have slopes $0.5114(2)$, $0.6648(7)$, $0.6669(6)$, $0.6657(6)$ in (a) from up to down and $0.501(1)$ in (b). We use the theoretic values of $r_{s0}$ and $M_{v0}$ as they equal almost the corresponding values obtained by fitting the raw data instead of the subtracted ones. Data from cooling at $H=0$ are obtained by cooling from $r=0$ with an initial order parameter of $10^{-10}$. It is found that the smaller the initial order parameter, the closer the slopes to the theoretical values.}
\label{rMv}
\end{figure}
One can also represent the transition by a susceptibility $d M/d r$, which exhibits a valley. The dependence of the position of its nadir, $r_v$, on $R$ is depicted in Fig.~\ref{rMv}(a). We find $r_v$ follows also Eq.~(\ref{rscale}) with again $\Upsilon=2/3$ and $r_0=r_{s0}$ for small rates. However, for the order parameter at $r_v$, viz. $M_v$, the slopes are not $1/3$ as seen in Fig.~\ref{rMv}(b) and can be even close to $1$ for vanishingly small rates.

The reason that we called this Thermal Class I as field-like class can be seen from
Eq.~(\ref{mnew}). For sufficiently small $R$ and thus $\tau$, $m$ is vanishingly small. As a
result, the dominant driving force is the effective field $K$ that is
equivalent to the field in the field driving case~\cite{zhongchen,Zhong2}.
From the view of Eq.~(\ref{Langevin 2}), a small $R$ means a small $\tau$ and thus
the first linear term is dominant and the second quadratic term can be ignored
in Eq.~(\ref{Kbar}). Consequently, a linearly varying $\tau$ is equivalent to
a linearly varying $\overline{K}$, which is just the field that drives the transition.

In this limit, Eq.~(\ref{mnew}) in the absence of the gradient can be solved analytically as it is a kind of Riccati equation~\cite{crc}. The result is~\cite{jung,Zhong2}
\begin{equation}\label{mas}
m(t)=-\sqrt[3]{\frac{4M_{s0}R}{v^2\lambda}}\frac{c_1Ai'\left(-\sqrt[3]{cR}t\right)+ c_2Bi'\left(-\sqrt[3]{cR}t\right)} {c_1Ai\left(-\sqrt[3]{cR}t\right)+c_2Bi\left(-\sqrt[3]{cR}t\right)},
\end{equation}
where $Ai$ and $Bi$ are the Airy functions, $c\equiv v\lambda^2M_{s0}/2$, $c_1$ and $c_2$ are constants to be determined by initial conditions, and a prime indicates a derivative with respect to the argument in this subsection. Therefore, one recovers Eq.~(\ref{rscale}) with $a_1=x_t/\sqrt[3]{g\lambda^2/2}$ and $r_0=r_{s0}$ from Eq.~(\ref{mas}) at $\sqrt[3]{v\lambda^2R/2}t=x_t$ at which $m=0$ and Eq.~(\ref{mscale}) with $a_2=\sqrt[3]{4/g^2\lambda}[c_1Ai'(0)+c_2Bi'(0)]/[c_1Ai(0)+c_2Bi(0)]$ and $M_0=M_{s0}$ from Eq.~(\ref{mas}) at $K=Rt=0$.

From Fig.~\ref{rMv}, one sees that although the valley temperature $r_v$ converges to $r_{s0}$ when $R\rightarrow0$ as $r_t$ does, $M_v$ does not tend to $M_{s0}$ but instead to $M_{v0}$ in the same limit, which is given by
\begin{equation}
M_{v0}=M_{s0}-2v/g.\label{mv0}
\end{equation}
This can be seen as follows.

We first notice that $r_v$ can only approach $r_{s0}$ for $R\rightarrow0$. This can be inferred from the inset of Fig.~\ref{rMs}. As $R\rightarrow0$, the transition between the two phases tends to $r_{s0}$. When $R=0$, the transition takes place vertically at $r_{s0}$. Accordingly, the valleys of $dM/dr$ approach $r_{s0}$ too. Also, $dM/dr$ itself diverges there. From Eq.~(\ref{mnew}) with the neglected cubic term reinstated, one finds at the nadir of the valley,
\begin{equation}
\frac{dm}{dr}=-\frac{m_{v}+M_{s0}}{\tau+vm_{v}+gm_{v}^2/2},
\end{equation}
which means either $m_{v0}=0$ or $m_{v0}=-2v/g$ at $R=0$ at which $\tau=0$ and $dM/dr=\infty$ but $m_v$ finite. By Eq.~(\ref{mshift}), the solution $m_{v0}=0$ corresponds to $M=M_{s0}$ and $m_{v0}=-2v/g$ to Eq.~(\ref{mv0}).

As $M_v$ approaches $M_{v0}$ at $R=0$, we can expand $M$ at $M_{v0}$, i.e., $M=M_{v0}+\overline{m}$, resulting in
\begin{equation}
\frac{\partial \overline{m}}{\partial t}=-\lambda\left( \tau \overline{m}-{\nabla}^{2}{\overline{m}-}\frac
{1}{2!}v\overline{m}^{2}-K + K_0\right)  , \label{mpnew}%
\end{equation}
from Eq.~(\ref{Me}), where $K_0\equiv2v^3/3g^2$. The differences between Eq.~(\ref{mpnew}) and Eq.~(\ref{mnew}) are only the new constant $K_0$ term and the opposite sign of the quadratic term. It is this opposite sign that renders $M>M_{v0}$ for both heating and cooling.
The solution of Eq.~(\ref{mpnew}) is again Eq.~(\ref{mas}) only with $-t$ replaced by $t+K_0/M_{s0}R$. This might be argued to be a shift of $r_{s0}$ to $r_{s0}-K_0/M_{s0}$. However, as pointed out above, $r_v$ converges to $r_{s0}$ instead of this shifted one.

We can now explain the results for small rates in Fig.~\ref{rMv}. On the one hand, as $r_v$ converges to $r_{s0}$, we must use Eqs.~(\ref{mnew}) and (\ref{mas}). The nadir of the valley must appear at a certain $x_v$ instead of $x_t$ at the transition temperature. This then results again in a $\Upsilon=2/3$ for small rates. On the other hand, we may directly use Eq.~(\ref{mpnew}) for $M_v$. But now the argument of the Airy functions become $x_v+(K_0\sqrt[3]{c}/M_{s0})R^{-2/3}$, which is strongly $R$ dependent for small $R$. Accordingly, $M_v$ is not just controlled by the prefactor $R^{1/3}$.

\subsubsection{\label{n12}Thermal Classes II and III}

Detailed analysis, however, reveals that there is another class with a
different $\Upsilon$. This can be readily seen from the cooling transition taking
place at $H=0$. In this case, the instability point is $(r_{s0}
^{-},M_{s0}^{-})=(0,0)$. As mentioned above, Eq.~(\ref{mnew}) and Eq.~(\ref{Me}) are then
identical up to the quadratic term. As the transition is at $r=0$, the leading
term of the free energy is again $\phi^{3}$ and ought to be described by a $\phi^{3}$ theory too.
Indeed, one finds that in this case $r_{v}$ and $M_v$ follow Eqs.~(\ref{rscale}) and (\ref{mscale}) with $\Upsilon$ and $\Upsilon_{m}$, now denoted as
$\ddot{\Upsilon}$ and $\ddot{\Upsilon}_{m}$, respectively, for Thermal Class III, of about $1/2$ distinct from the above case, as can be seen from Fig.~\ref{rMv}. Accordingly, it is
important to use Eq.~(\ref{rscale}) instead of Eq.~(\ref{a}) to correctly
identify the scaling at least in this special case in which the heating and the cooling transition behave distinctly.

In fact, similar to the purely thermal class of the cooling transition at $H=0$, there exist other cases in which the hysteresis exponents are close to $1/2$. Figure~\ref{rMv} demonstrates that for a range of large $R$ values, $\Upsilon$ and $\Upsilon_{m}$ are close also to $1/2$. Moreover, even in the heating
transition, they can also be close to $1/2$ as Fig.~\ref{rMv} indicates. We classify these as Thermal Class II and denote its $\Upsilon$ and $\Upsilon_{m}$ with an overdot. In Fig.~\ref{rMs}(a), the slopes for $r_t$ also decrease for large $R$. However, $M_t$ appears to be saturated at high rates. Although, in mean field, the hysteresis exponents of this class are identical with those of Thermal Class III, the above results indicate that this class emerges when $\tau m$ instead of $K$ dominates. In other words, it is
described by Eq.~(\ref{Langevin 3}) with a linearly varying $\overline{\tau}$ as
the second term in Eq.~(\ref{taubar}) can be ignored for large $\tau$. As it can cross over to the field-like class (see below), we also call Thermal Class II as partly thermal class in order to distinguish it from the purely thermal class.

Indeed, when $K$ is omitted, Eq.~(\ref{mnew}) in the absence of the gradient becomes a Bernoulli's equation that is solved by~\cite{crc}
\begin{equation}
m=\frac{\sqrt{R}\exp\left\{-\frac{1}{2}\lambda\left[\left(\frac{\tau}{\sqrt{R}}\right)^2- \left(\frac{\tau_0}{\sqrt{R}}\right)^2\right]\right\}}{\sqrt{R}m_0+\frac{1}{2}\lambda v \int_{\tau_0/\sqrt{R}}^{\tau/\sqrt{R}}\exp\left[-\frac{1}{2}\lambda(\tau'^2-\tau_0^2/R)\right]d\tau'},
\end{equation}
where $m(\tau_0)\equiv m_0$ and the integral can be expressed as Error functions. One sees that $\tau$ appears in the solution through the combination $\tau/\sqrt{R}$. Consequently, at certain loci of certain functions of the solution, $\tau$ is expected to be proportional to $R^{1/2}$ though with possible corrections from the initial condition.

\subsubsection{\label{crossover}Crossover between Thermal Classes I and II}
We have seen that except cooling in the absence of an external field, a situation which leads to the purely thermal class Thermal Class III, in the presence of an external field, there exist two thermal classes I and II that exhibit different behaviors for large and small rates, respectively.

In the median range of $R$ between the two extremal cases, one finds from Fig.~\ref{rMv} that $\Upsilon$ falls between $2/3$ and
$1/2$. This stems from the nonlinearity of $\overline{K}$ and $\overline{\tau}$ and
reflects the competition between their two ingredients in Eqs.~(\ref{Kbar}) and
(\ref{taubar}), respectively.

Owing to this crossover, only sufficiently small $R$ can give rise to
$\Upsilon=2/3$. However, it was found numerically and
even experimentally~\cite{Yildiz} that the exponent of the thermal hysteresis
areas in the $M^{2}$-$r$ frame is still close to $2/3$ for not-so-large $R$. The
area in the $M$-$r$ frame is
\begin{align}
A  &  =\oint Mdr=\int_{r_{\rm{down}}}^{r_{\rm{up}}}(M_{+}%
-M_{-})dr\nonumber\\
&  \simeq\left(  \int_{r_{v}^{-}}^{r_{s0}^{-}}+\int_{r_{s0}^{-}}^{r_{s0}^{+}%
}+\int_{r_{s0}^{+}}^{r_{v}^{+}}\right)  (M_{s0}^{+}+m_{+}-M_{s0}^{-}%
-m_{-})dr\nonumber\\
&  \sim A_{0}+\left(M_{s0}^{+}-M_{s0}^{-}\right)\left(a_{1}^{+}R^{\Upsilon^{+}}+a_{1}^{-}R^{\Upsilon^{-}}\right) \label{aint}%
\end{align}
to the leading order in $R$, where ${+}$ (${-}$) denotes variables in heating (cooling),
$r_{\rm{down}}$ ($r_{\rm{up}}$) the lowest (highest) temperature
delimitating the hysteresis loop, $A_{0}=\int_{r_{s0}^{-}}^{r_{s0}^{+}}(M_{+}-M_{-})dr$ at equilibrium, and uses have been made of Eqs.~(\ref{mshift}) and
(\ref{rscale}). We have neglected the contributions from Eq.~(\ref{mscale}) since $m_t=M_t-M_{s0}$ is vanishingly small as can be seen from the inset of Fig.~\ref{rMs}. Indeed, we could not detect this leading order in $R$ for very small $R$. Note that we have employed $r_v$ for $r$ but $M_t$ for $M$, as $r_v$ extends beyond $r_t$ both in heating and in cooling and may provide a better approximation than the latter does but $M_v$ does not converge to $M_{s0}$. A similar approximate expression can be written for the area in the $M^{2}$-$r$ frame, which is proportional to the energy dissipation. We see
that both areas show similar behavior in the approximations embodied in Eq.~(\ref{aint}). However, the
exponent found from the $M^{2}$-$r$ frame is constantly bigger than that from the $M$-$r$ frame no matter whether $M_{s0}$ is large or small. One possible reason is that the approximation near the heating spinodal is always better in the former frame no matter whether $M_{s0}$ is large or small, while the approximation near the cooling spinodal is always poor and may obscure the behavior there, as can also been seen from the inset of Fig.~\ref{rMs}. Comparing Fig.~\ref{rMs}(a) with Fig.~\ref{rMv}(a), one finds that the fitted ranges of $R$ are wider and the slopes are larger for $r_v$ than those for $r_t$ in heating. This may indicate that the area may have even favorable results and may thus be another reason.

\section{\label{scaling} Scaling theory}

In order to relate the hysteresis exponents $\Upsilon$ and $\Upsilon_{m}$ to more
fundamental exponents, We now perform a scaling analysis for the models.

\subsection{\label{gsf}General scaling forms}

First we consider the massless theory~(\ref{Langevin 2}). Following
\cite{zhongchen}, we make a scale transformation to it by reducing the length
scale by $\rho$, i.e., $|\mathbf{r^{\prime}}|=|\mathbf{r}|/\rho$. Then,
\begin{align}
\nabla^{\prime2}  &  =\rho^{2}\nabla^{2},\qquad(\lambda t)^{\prime} =(\lambda
t)\rho^{-z},\qquad\varphi^{\prime}=\varphi\rho^{\beta/\nu
},\nonumber\\
\ \overline{K}^{\prime}  &  =\overline{K}\rho^{\beta\delta/\nu},\qquad v^{\prime} =v%
\rho^{y},\qquad\tau^{\prime}=\tau\rho^{1/\nu}, \label{m1}%
\end{align}
where $\beta$, $\nu$, $\delta$, and $z$ are the instability exponents
corresponding to the standard critical exponents, and $y$ is a constant. We
have set $\lambda$ as the unit of time $t$ as usual \cite{zhongchen}. Note
that $M_{s0}$ changes also like $\varphi$. Invariance of Eqs.~(\ref{Langevin 2}%
) and (\ref{Kbar}) under this transformation leads to $z=1/\nu=2$,
$\beta(\delta-1)=1$, $y=(1-\beta)/\nu$, and
\begin{equation}
\varphi(t,\overline{K})=\rho^{-\beta/\nu}\varphi(t\rho^{-z},\overline{K}\rho^{\beta
\delta/\nu}), \label{mrho}%
\end{equation}
or
\begin{equation}
\varphi(t,\overline{K})=t^{-\beta/\nu z}f_{K}(\overline{K}t^{\beta\delta/\nu z})
\label{mrr}%
\end{equation}
by choosing such a scale $\rho$ that $t\rho^{-z}$ becomes a constant, where
$f_{K}$ is a scaling function. We have neglected dimensional factors for simplicity.

Note that we can also obtain the scaling laws among the exponents from the
invariance of Eq.~(\ref{mnew}). The only problem with this dynamic equation is
the redundancy of the parameters.

We have suppressed the coupling $v$ in Eqs.~(\ref{mrho}) and (\ref{mrr}). In
fact, for a renormalizable theory~\cite{Justin,amitb,Kleinert,Vasilev}, $v$
should be dimensionless and thus should keep invariant. This dictates $y=0$ or
$\beta=1$ and hence $\delta=2$. In addition, the susceptibility exponent
$\gamma=\beta(\delta-1)=1$. The scaling laws $\gamma/\nu=2-\eta$ and
$\alpha+2\beta+\gamma=2$ then lead to $\eta=0$ and $\alpha=-1$, which in turn
results in an upper critical dimension $d_{c}=6$ owing to the hyperscaling law
$\alpha=2-d\nu$, in agreement with the dimensional
analysis~\cite{zhongchen,Zhong2}. These complete the list of the usual
mean-field instability exponents. They all share identical meaning with their critical counterparts~\cite{zhongchen,Zhong2}.

Equation~(\ref{mrr}) can be written in an FTS form in terms of
the rate $R$. In linearly temperature-driven FOPTs,
$\tau=r-r_{s0}=Rt$. Using Eq.~(\ref{Kbar}), we obtain
\begin{equation}
\varphi=(\tau/R)^{-\frac{\beta}{\nu z}}f_{K}\left[  \left(  {\tau^{\frac{2\nu
z+\beta\delta}{\nu z}}/2v-M_{s0}\tau^{\frac{\beta\delta+\nu z}{\nu z}}}\right)
{R}^{-\frac{\beta\delta}{\nu z}} \right]  . \label{solution 3}%
\end{equation}
Therefore, from Eq.~(\ref{shift}),
\begin{align}
m(\tau,R) =  &  -\tau/{v}+(\tau/R)^{-\frac{\beta}{\nu z}} \times\nonumber\\
&  f_{K}\left[  \left(  {\tau^{\frac{2\nu z+\beta\delta}{\nu z}}/2v-M_{s0}%
\tau^{\frac{\beta\delta+\nu z}{\nu z}}}\right)  {R}^{-\frac{\beta\delta}{\nu
z}}\right]  . \label{solution 4}%
\end{align}

For the purely massive theory~(\ref{Langevin 3}), although $\overline{\tau}$ is like $\tau$, we assume that it may scale distinctly from the latter with an exponent $\overline{\nu}$. So,
\begin{equation}
\varphi(t,\overline{\tau})=\rho^{-\beta/\nu}\varphi(t\rho^{-z},\overline{\tau
}\rho^{1/\overline{\nu}}). \label{mrt}%
\end{equation}
Then, similar method leads to
\begin{align}
m(\tau,R)=  &  (\tau/R)^{-\frac{\beta}{\nu z}}f_{\overline{\tau}}\left[
\left(  \tau^{\frac{2\overline{\nu}z+2}{\overline{\nu}z}}-2vM_{s0}\tau^{\frac{\overline{\nu
}z+2}{\overline{\nu}z}}\right)  {R}^{-\frac{2}{\overline{\nu}z}}\right] \nonumber\\
&  -\tau/v-\sqrt{\tau^{2}-2vM_{s0}\tau)}/v \label{solution 5}%
\end{align}
using Eqs.~(\ref{shift}) and (\ref{taubar}), where ${f}_{\overline{\tau}}$ is another
scaling function.

Of course, for the purely thermal class, $M_{s0}=0$ and $\overline{\tau}$ is just $\tau$. So, the scaling form~(\ref{solution 5}) simplifies to
\begin{equation}
M(\tau,R)=   (\tau/R)^{-\frac{\beta}{\nu z}}f_{\tau}
\left(  \tau^{\frac{1+\nu z}{\nu z}} {R}^{-\frac{1}{\nu z}}\right),\label{solution 6}%
\end{equation}
or a standard FTS form
\begin{equation}
M(\tau,R)=   R^{\frac{\beta}{1+\nu z}}f_{\tau1}
\left(  \tau{R}^{-\frac{1}{1+\nu z}}\right),\label{solution 6r}%
\end{equation}
where $f_{\tau}$ is yet another scaling function and $f_{\tau1}(x)=x^{-\beta/\nu z}f_{\tau}(x^{1+1/\nu z})$.

In mean field, the value of $\overline{\nu}$ is identical with $\nu$. However, we shall see that they are different in value in non-mean-field case.

\subsection{Limiting cases and their thermal hysteresis exponents}

Equations~(\ref{solution 4}) and (\ref{solution 5}) are equivalent general
descriptions of the scaling behavior in the vicinity of the instability points
of the temperature-driven FOPTs. As they are quite complicate owing to the
nonlinearity of $\overline{K}$ and $\overline{\tau}$, the thermal hysteresis exponents
cannot be read out directly. Note that if one varied $\overline{K}$ and $\overline{\tau
}$ instead of $\tau$ itself linearly, simple FTS form would be
present, similar to Eq.~(\ref{solution 6}). Further, the two parameters, $\overline{K}$ and $\overline{\tau} $, in fact,
contain only one independent combination, $\tau^{2}-2vM_{s0}\tau$. So, simple
scaling forms would be obtained by varying this factor linearly. However,
simple scaling forms do appear and the hysteresis exponents can be readily
identified in the two limits in which either the linear or the quadratic term
alone in $\overline{K}$ and $\overline{\tau}$ is kept. This results in the field-like
universality class and the partly thermal class, respectively. On the other hand, the scaling forms~(\ref{solution 6}) or (\ref{solution 6r}) gives rise to the purely thermal class.

\subsubsection{Thermal Class I}

The field-like universality class can be obtained either from
Eq.~(\ref{solution 4}) or (\ref{solution 5}). We start with the
massless theory, Eq.~(\ref{solution 4}).

When the term linear in $\tau$ in $\overline{K}$ dominates due to a small $R$,
we omit the quadratic term in Eq.~(\ref{solution 4}) and obtain
\begin{equation}
m(\tau,R) =R^{\frac{\beta}{\beta\delta+\nu z}}f_{K1}\left(  \tau{R}%
^{-\frac{\beta\delta}{\beta\delta+\nu z}}\right)  -\tau/{v} \label{mrk1}%
\end{equation}
with $f_{K1}(x)=x^{-\beta/\nu z}f_{K}[-M_{s0}x^{(\beta\delta+\nu z)/\nu z}]$.

From Eq.~(\ref{mrk1}), thermal hysteresis exponents can be obtained. Indeed,
at $\tau=0$ or $r=r_{s0}$, Eq.~(\ref{mrk1}) results in Eq.~(\ref{mscale}) with
\begin{equation}
\Upsilon_{m}=\frac{\beta}{\beta\delta+\nu z}, \label{nTm}%
\end{equation}
$M_{s0}=M_{0}$, and $a_{2}=f_{K1}(0)$ by using Eq.~(\ref{mshift}). At $M=M_{s0}$ or
$m=0$ at which $\tau=\tau_{t}$, we have
\begin{equation}
f_{K1}\left(  \tau_{t}{R}^{-\frac{\beta\delta}{\beta\delta+\nu z}}\right)
=\tau_{t} R^{\frac{-\beta}{\beta\delta+\nu z}}/{v}. \label{taut1}%
\end{equation}
Since $\delta>1$, the argument on the left dominates. So,
\begin{equation}
\tau_{t}=r_{t}-r_{s0}=R^{\Upsilon}f_{K2}(R^{\Upsilon-\Upsilon_{m}}), \label{tautk}%
\end{equation}
where $f_{K2} ^{-1}(x)=vf_{K1}(x)/x$,
\begin{equation}
\Upsilon=\frac{\beta\delta}{\beta\delta+\nu z}, \label{nT}%
\end{equation}
and $r_{0}=r_{s0}$ from Eqs.~(\ref{rscale}) and (\ref{tau}). Note that
$\Upsilon-\Upsilon_{m}=\gamma/(\beta\delta+\nu z)>0$ and thus $f_{K2}(x)$ constitutes
a correction to the leading scaling correctly~\cite{Wegner}.

On the other hand, from the scaling form of the purely massive theory,
Eq.~(\ref{solution 5}), keeping only the leading $\tau$ term, we have
\begin{equation}
m(\tau,R)=R^{\frac{\beta\overline{\nu}}{(2+\overline{\nu}z)\nu}}f_{\overline{\tau}1}\left(  \tau{R}%
^{-\frac{2}{2+\overline{\nu}z}}\right)  +\sqrt{-2M_{s0}\tau/v} \label{mrt1}%
\end{equation}
with $f_{\overline{\tau}1}(x)=x^{-\beta/\nu z}f_{\overline{\tau}}[-2vM_{s0}x^{(2+\overline{\nu
}z)/\overline{\nu}z}]$. So, at $\tau=0$, we again arrive at Eq.~(\ref{mscale}) but
with
\begin{equation}
\Upsilon_{m}=\frac{\beta\overline{\nu}}{(2+\overline{\nu}z)\nu} \label{nTmt}%
\end{equation}
and $a_{2}=f_{\overline{\tau}1}(0)$ now. At $m=0$, we write
\begin{equation}
\tau_{t}=r_{t}-r_{s0}=R^{\Upsilon}f_{\overline{\tau}2}(R^{\Upsilon/2-\Upsilon_{m}}) \label{tautt}%
\end{equation}
with
\begin{equation}
\Upsilon=\frac{2}{2+\overline{\nu}z}, \label{nTt}%
\end{equation}
where $f_{\overline{\tau}2}^{-1}(x)=\sqrt{-v/2M_{s0}}f_{\overline{\tau}1}(x)/\sqrt{x}$. In this
case, however, since $\beta=1$, $\Upsilon/2-\Upsilon_{m}=0$. So, the correction
$f_{\tau2}$ in Eq.~(\ref{tautt}) is just a constant.

In mean field, we have $\overline{\nu} z=1$ and $\beta\delta=2$. So, both
Eqs.~(\ref{nTm}) and (\ref{nTmt}) and both Eqs.~(\ref{nT}) and (\ref{nTt})
yield $\Upsilon_{m}=1/3$ and $\Upsilon=2/3$, respectively, consistently in agreement
with numerical results, though, as mentioned, the purely massive theory may be real only in the cooling transitions in the absence of the external field.

We have shown that the two reduced theories are indeed equivalent in mean field. In
fact, since we consider essentially the effective field $K$ in this class and
ignore the mass term, we can only study the massless theory with $K$ replacing
$\overline{K}$ directly. This has been studied previously~\cite{zhongchen,Zhong2} and
yields of course identical exponents albeit without the correction $f_{K2}$.

In the case of $\overline{m}$ in Eq.~(\ref{mpnew}), $K_0$ acts as a relevant constant external field. Accordingly, the scaling function, $f_{K1}$ for example, has $K_0R^{-\beta\delta/(\beta\delta+\nu z)}$ as its additional argument. This destroys the simple scaling in consistence with the mean-field theory and Fig.~\ref{rMv}(b).

\subsubsection{Thermal Class II}

In this case, we keep only the quadratic terms in $\overline{K}$ and $\overline{\tau}$.
This means to neglect the effective field $K$ and hence $\overline{\tau}$ is just
$\tau$ and $\varphi$ is just $m$. Accordingly, for the purely massive theory, Eq.~(\ref{solution 5}) becomes
\begin{equation}
m(\tau,R) =R^{\frac{\beta\overline{\nu}}{(1+\overline{\nu} z)\nu}}f_{\overline{\tau}3}\left(  \tau
{R}^{-\frac{1}{1+\overline{\nu} z}}\right)  \label{mrt2}%
\end{equation}
with $f_{\overline{\tau}3}(x)=x^{-\beta/\nu z}f_{\overline{\tau}}[x^{(2+2\overline{\nu}
z)/\overline{\nu} z}]$. Thus, one obtains readily
\begin{equation}
\dot{\Upsilon}_{m}    =\frac{\beta\overline{\nu}}{(1+\overline{\nu} z)\nu},\qquad
\dot{\Upsilon}    =\frac{1}{1+\overline{\nu} z}. \label{nTt1}%
\end{equation}

The scaling form of the massless theory, Eq.~(\ref{solution 4}), can also lead to this class, of course. It
becomes
\begin{equation}
m(\tau,R) =R^{\dot{\Upsilon}_{m}}f_{K3}\left(  \tau{R}^{-\dot{\Upsilon}}\right)
-\tau/{v} \label{mrk2}%
\end{equation}
with $f_{K3}(x)=x^{-\beta/\nu z}f_{K}[x^{(\beta\delta+2\nu z)/\nu z}/v]$ and
\begin{equation}
\dot{\Upsilon}_{m}    =\frac{2\beta}{\beta\delta+2\nu z},\qquad
\dot{\Upsilon}    =\frac{\beta\delta}{\beta\delta+2\nu z}, \label{nTtk1}%
\end{equation}
in the same limit. However, there is a subtlety here. In mean-field
approximation, $\delta=2$ and hence $\dot{\Upsilon}_{m}=\dot{\Upsilon}$ and no
correction is necessary. In $d<d_{c}$, $\delta<2$~\cite{Zhong2} and hence $\dot{\Upsilon}_{m}>\dot{\Upsilon}$. The argument from Eq.~(\ref{taut1}) to (\ref{tautk}) then gives rise to a surprising consequence:
both the temperature hysteresis exponent and the
order-parameter hysteresis exponent equal $\dot{\Upsilon}_{m}$ and the correction exponent is $\dot{\Upsilon}_{m}-\dot{\Upsilon}>0$.

In mean field, both theories again yield the same exponents of $\dot{\Upsilon}_{m}=1/2=\dot{\Upsilon}$ in agreement with numerical results shown in
Fig.~\ref{rMv}.

\subsubsection{Thermal Class III}
From similar procedures, we obtain directly the hysteresis exponents for this purely thermal class as
\begin{equation}
\ddot{\Upsilon}_{m}    =\frac{\beta}{1+\nu z},\qquad
\ddot{\Upsilon}    =\frac{1}{1+\nu z}, \label{nTt3}%
\end{equation}
using the scaling form~(\ref{solution 6r}). In the mean-field approximation, $\overline{\nu}=\nu$ and thus Eq.~(\ref{nTt3}) is just Eq.~(\ref{nTt1}).

\subsection{\label{gptc}General purely thermal class}
We mentioned above that the cooling transition of a $\phi^6$ model at $H=0$ falls into a different class. In this section, we study briefly this issue.

Consider the general model~(\ref{Langevin 4}) with an external field $H$. Although in this class, $H=0$, we still keep it in order to gather sufficient information. A scaling transformation similar to Eq.~(\ref{m1}) with $y=0$ then leads to
$z=1/\nu=2$ and $\beta=1/(\sigma-1)$. Combining with the scaling laws mentioned in Sec.~\ref{gsf}, we find then $\gamma=1$, $\delta=\sigma$, $\eta=0$, $\alpha=(\sigma-3)/(\sigma-1)$, and the upper critical dimension $d_c=2(\sigma+1)/(\sigma-1)$. This $d_c$ conforms again with na\"{\i}ve dimensional analysis. Therefore, from Eq.~(\ref{nTt3}), we see that in this case, although the value of $\ddot{\Upsilon}$ is identical, that of $\ddot{\Upsilon}_{m}$ may be different.

For the $\phi^6$ model, $\sigma=3$. So, we have $\beta=1/2$, $d_c=4$, and so on, though $z=1/\nu=2$. In fact, this is just the usual $\phi^4$ model for the critical phenomena of the Ising universality class. The hysteresis exponents for this class have already been computed in Ref.~\cite{zhonge}. They are different from those of the thermal transitions of the $\phi^3$ model studied here, which has $\sigma=2$. However, we still class both models into Thermal Class III as both concern with the cooling transition in the absence of the external field.

\section{\label{rgt}RENORMALIZATION GROUP THEORY}

Our task is to confirm the scaling for $d<d_{c}$ and to calculate the scaling exponents.
This can be done by an RG
analysis~\cite{Wilson,Ma,Justin,amitb,Kleinert,Vasilev}. An RG study of the
dynamics of the full $\phi^{3}$ theory~(\ref{mnew}) with the Gaussian white
noise~(\ref{noise}) for the Yang-Lee edge singularity has been presented
in~\cite{Breuer} using the shift invariance. A detailed RG study of the
massless theory~(\ref{Langevin 2}) has been performed in~\cite{Zhong2} with
hysteresis exponents for the field-driven FOPTs computed in three- and two-loop
orders for the static and dynamic exponents, respectively. Accordingly, we may just insert those exponents for the relevant classes.
In fact, the exponents for the field-like class are just identical to the
field-driven FOPTs. As the massless and massive theories are equivalent,
comparing Eqs.~(\ref{nTm}) and (\ref{nT}) with Eqs.~(\ref{nTmt}) and
(\ref{nTt}) and Eq.~(\ref{nTt1}) with (\ref{nTtk1}), one sees that for both pairs of exponents to be identical, one
must have surprisingly $\overline{\nu}=2\nu/\beta\delta$, whose values are indeed equal in mean field. We shall study the RG theory to prove it and the equivalence of the two theories.

\subsection{\label{fluctuations}Fluctuation shifts and relationship between temperature and field}
We start with the full dynamic model~(\ref{f}) with the Gaussian white
noise~(\ref{noise}). It is well established that it can be recast into a
field-theoretic form with a dynamic functional
\cite{Janssen79,janssen,Tauber,Justin,Vasilev},
\begin{align}
I[{\phi,\widetilde{\phi}}]=   \int d\mathrm{\mathbf{x}}dt  &\left\{  {\widetilde{\phi}%
}\left[  \frac{\partial{\phi}}{\partial t}+\lambda(r-{\nabla}^{2}){\phi
}\right.  \right.  \nonumber\\
&  \left.  \left.  ~~{+}\frac{1}{2!}\lambda{w\phi}^{2}+\frac{1}{3!}\lambda
{g\phi}^{3}-\lambda H\right]  {-\lambda\widetilde{\phi}}^{2}\right\}
,\label{action}%
\end{align}
by introducing an auxiliary response field $\widetilde{\phi}$~\cite{martin}.
Then the generating functionals for connected and vertex response and
correlation functions can be defined and perturbation expansions using Feynman
diagrams can be set up.

Assuming that the instability point is now at $(r_s, M_s)$ and $\widetilde{\phi}=0$, we expand at $M_s$ but near $r_s$ for a constant uniform field $H$ by
\begin{equation}
{\phi}=M_s+\varphi,\qquad \widetilde{\phi}=\widetilde{\varphi}, \label{mshift 2}%
\end{equation}
and neglect again the cubic term. Equation~(\ref{action}) then changes into
\begin{align}
I[{\varphi,\widetilde{\varphi}}]=   \int d\mathrm{\mathbf{x}}dt & \left\{  {\widetilde{\varphi}%
}\left[  \frac{\partial{\varphi}}{\partial t}+\lambda(\tau-{\nabla}^{2}){\varphi
}\right.  \right.  \nonumber\\
&  \left.  \left.  ~~{+}\frac{1}{2!}\lambda{v\varphi}^{2}-\lambda K\right]  {-\lambda\widetilde{\varphi}}^{2}\right\}
,\label{action2}%
\end{align}
with
\begin{subequations}
\label{tauc}%
\begin{align}
\tau &  =r+wM_s+\frac{1}{2}gM_s^{2},\label{tau 3}\\
K  &  =H-rM_s-\frac{1}{2}wM_s^{2}-\frac{1}{3!}gM_s^{3},\label{tau 4}\\
v  &  =w+gM_s, \label{g 1}%
\end{align}
\end{subequations}
similar to Eq.~(\ref{taug}). In the mean-field approximation, $r_s=r_{s0}$ and $M_s=M_{s0}$ and thus we recover Eq.~(\ref{taug}). When fluctuations are taken into account, the instability point now shifts from $\tau_{s0}=K_{s0}=0$ to $\tau=\tau_s$ and $K=K_s$ determined by~\cite{zhongchen,Zhong2}
\begin{subequations}
\begin{eqnarray}
\Gamma^{(11)}({\bf 0},0)&=&0,\label{taus}\\
\langle\varphi\rangle=0,\quad {\rm ~or,}&&\Gamma^{(10)}({\bf 0},0)=\lambda K_s,\label{hs}
\end{eqnarray}
\end{subequations}
i.e.,
\begin{subequations}
\label{tauc3}%
\begin{align}
\tau_{s} &  =\frac{1}{2} v^{2}\int \frac{d{\mathbf{k}}}{(2\pi)^d}\frac{1}{({\mathbf{k}}^{2}+\tau_{s})^{2}%
},\label{fluctuation5}\\
K_{s}&  =\frac{1}{2} v\int \frac{d{\mathbf{k}}}{(2\pi)^d}\frac{1}{{\mathbf{k}}^{2}+\tau_{s}%
},\label{fluctuation6}%
\end{align}
\end{subequations}
to one-loop order, or
\begin{subequations}
\label{tauc4}
\begin{align}
\tau_{s}^{3-d/2} &  =\frac{\Gamma(3-d/2)\Gamma(d/2)}{2(4-d)}%
v^{2}N_{d},\label{fluctuation7}\\
K_{s} &  =\frac{\Gamma(3-d/2)\Gamma(d/2)}{(2-d)(4-d)}vN_{d}\tau
_{s}^{d/2-1},\label{fluctuation8}%
\end{align}
\end{subequations}
where $\Gamma^{(\widetilde{N}N)}$ is the vertex function, $N_{d}=2/[(4\pi)^{d/2}\Gamma(d/2)]$ ($\Gamma$ is the Euler gamma function), the angle brackets denote the average over the action, and $\{{\mathbf{q}},\omega\}$ represents a set of momenta and frequencies. Similar to the theory of critical phenomena~\cite{Justin,amitb,Kleinert,Vasilev}, Eq.~(\ref{taus}) implies the divergence of the inverse susceptibility at the instability point. Divergences of the correlation function at long distances and the correlation length at the instability point can also be derived within the theory. However, the real correlation length of the original FOPT does not diverge because of the irrelevant higher order terms. Equation~(\ref{tauc4}) satisfies a relation
\begin{equation}
K_s+\tau_s^2/2v=0\label{kctc}
\end{equation}
near $d=6$.
Using $\tau_s$ and $K_s$, which satisfy Eq.~(\ref{tauc}) at $r_s$, we can rewrite the first two equations in~(\ref{tauc}) as
\begin{subequations}
\label{tauc1}
\begin{align}
\tau-\tau_s &  =r-r_s,\label{tau rc}\\
K-K_s  &  =-M_s(r-r_s)=-M_s(\tau-\tau_s).\label{tau kc}
\end{align}
\end{subequations}
This indicates that the relation between $\tau$ and $K$ for temperature-driven FOPTs is still valid after the shifts even if the fluctuations are taken into account.

One sees that the cooling instability point in the absence of the external field now has generally $M_s\neq0$ because $K_s$ is finite for a $\varphi^3$ theory. This is qualitatively different from a $\varphi^4$ $\sigma=3$ theory, in which the symmetry forbids a $K_s$ term. Therefore, although in mean-field we have Thermal Class III described by the purely thermal model, it is removed by fluctuations. Yet, it survives in the generalized purely thermal model with an odd $\sigma$.

\subsection{Renormalization factors and their relations}
After the mass and field renormalizations, the theory becomes massless at the instability point. Consequently, its perturbation expansions are plagued with infrared divergences. Yet, it turns out that these
divergences can be removed by the renormalization factors $Z$ defined
as~\cite{Breuer,Vasilev}
\begin{align}
\varphi &  =Z_{\varphi}^{1/2}\varphi_{R},\qquad\widetilde{\varphi}=Z_{\widetilde
{\varphi}}^{1/2}\widetilde{\varphi}_{R},\qquad\tau-\tau_{s}=Z_{{\varphi}}^{-1}Z_{\tau}\tau_{R}%
,\nonumber\\
\lambda &  =Z_{\lambda}\lambda_{R}=\left(  Z_{{\varphi}}/Z_{\widetilde{\varphi}%
}\right)  ^{1/2}\lambda_{R},\qquad u=vN_{d}^{1/2}%
\mu^{-\epsilon/2},\nonumber\\
u &  =Z_{\varphi}^{-3/2}Z_{v}u_{R},\qquad v =Z_{\varphi}^{-3/2}Z_{v}v_{R},\nonumber\\
K&-K_{s}   =Z_{{\varphi}}^{-1/2}\left(  K_{R}+Z_{0}\tau_{R}^{2}/2v_{R}\right),\label{renormalization factors}%
\end{align}
such that the renormalized vertex function
\begin{equation}
\Gamma_{R}^{(\widetilde{N}N)}(\{{\mathbf{q}},\omega\})=Z_{\widetilde{\varphi}%
}^{\widetilde{N}/2}/Z_{{\varphi}}^{N/2}\Gamma^{(\widetilde{N}N)}%
(\{{\mathbf{q}},\omega\})\label{gnn}%
\end{equation}
(for $(\widetilde{N},N)\neq(1,0)$) in terms of the renormalized functions becomes finite, since the dimensional
poles at $\epsilon=6-d\rightarrow0$ have then been subtracted and just
subtracted in the minimal RG scheme with dimensional regulation \cite{hooft},
where the subscripts $R$ denote renormalized variables and $\mu$ is an arbitrary
momentum scale. This RG method has an additional advantage of decoupling
statics from dynamics \cite{Dominicis} so that the static renormalization
factors can be chosen as the equilibrium ones.

The renormalization factors are, however, not all independent.
The shift symmetry of Eq.~(\ref{action2}) with respect to the shift of the
order parameter gives rise to some exact relations among them~\cite{Breuer,Vasilev}.

For an arbitrary shift of ${\varphi}^{\prime}=\varphi+c$, due to the
invariance~\cite{Breuer,Vasilev}, the shifted variables, Eq.~(\ref{tkprimed}),
ought to share the same relations and renormalization factors to the original
ones. So,
\begin{equation}
\varphi^{\prime}=Z_{\varphi}^{1/2}\varphi_{R}^{\prime}={\varphi}%
+{c}=Z_{{\varphi}}^{1/2}{\varphi}_{R}+Z_{c}c_{R}, \label{factor 1}%
\end{equation}
which means $Z_{c}=Z_{\varphi}^{1/2}$ since $Z_{\varphi}^{-1/2}Z_{c}$ is a
finite quantity that in the minimal renormalization reduces to unity. Applying
this result with Eq.~(\ref{renormalization factors}) to Eq.~(\ref{tauprimed}),
we have
\begin{align}
\tau^{\prime}  &  =Z_{\varphi}^{-1}Z_{\tau}\tau_{R}^{\prime}+\tau_{s}=\tau-{v}c\label{factor 2}\\
&  =Z_{{\varphi}}^{-1}Z_{\tau}\tau_{R}-Z_{{\varphi}}^{-1}Z_{v}{v}_{R}%
c_{R}+\tau_{s},\nonumber
\end{align}
which leads to $Z_{\tau}=Z_{v}$. Similarly, from
Eqs.~(\ref{renormalization factors}) and (\ref{kprimed}) we get
\begin{equation}
{K}_{R}^{\prime}=K_{R}+(Z_{0}+Z_{\tau})\left(  \tau_{R}c_{R}-\frac{1}{2}%
v_{R}c_{R}^{2}\right)  . \label{factor 5}%
\end{equation}
Therefore, we have
\begin{equation}
Z_{\tau}=Z_{v}=1-Z_{0}. \label{field 6}%
\end{equation}
This indicates that among the four static renormalization factors introduced in
Eq.~(\ref{renormalization factors}), only two are independent. The $Z$ factor
for the vertex will be chosen to determine the fixed point. So, the other one
will determine the only independent static exponent. The two other factors,
$Z_{\lambda}$ and $Z_{\widetilde{\varphi}}$, which are related by the
fluctuation-dissipation theorem~\cite{Janssen79,janssen,Tauber,Justin,Vasilev}
to ensure a correct static limit~\cite{Zhong2}, determine one independent
dynamic exponent.

Equation~(\ref{field 6}) can also be obtained from Ward's identities stemming
from the continuous shift symmetry~\cite{Breuer}. Taking the derivative of the
shifted vertex functions with respect to $c$ at $c=0$ results in~\cite{Breuer}
\begin{align}
\Gamma_{R}^{(\widetilde{N}N+1)}(\{{\mathbf{q}},\omega\},\tau_{R})=  &
Z_{v}Z_{\tau}^{-1}v_{R}\frac{\partial}{\partial\tau_{R}}\Gamma_{R}%
^{(\widetilde{N}N)}(\{{\mathbf{q}},\omega\},\tau_{R})\nonumber\\
&  +\left(Z_{\tau}+Z_{0}Z_{v}Z_{\tau}^{-1}\right)\tau_{R}\delta_{N1}, \label{mass}%
\end{align}
which leads indeed to Eq.~(\ref{field 6}) since the renormalized vertex
functions possess no poles and the combinations of the $Z$ factors must be finite, where $\delta_{N1}$ is the Kronecker delta function.

To one-loop order, it is readily found~\cite{zhongchen,Zhong2}
\begin{equation}
Z_{{\varphi}}    =1-u_{R}^{2}/6\epsilon,\quad
Z_{v}    =1-u_{R}^{2}/\epsilon,\quad
Z_{\widetilde{\varphi}}    =1-u_{R}^{2}/3\epsilon. \label{zz}%
\end{equation}
These static and dynamic factors have been found to three-loop
order~\cite{Alcantara} and two-loop order~\cite{Breuer}, respectively.

It is instructive to know the renormalization of $\overline{K}$ and $\overline{\tau}$
which are $K^{\prime}$ and $\tau^{\prime}$ at some particular $c$. From
Eqs.~(\ref{Kbar}), (\ref{taubar}), (\ref{renormalization factors}), and
(\ref{field 6}), we find
\begin{align}
\overline{K}  &  =(K-K_{s})+(\tau-\tau_{s})^{2}/2v\nonumber\\
&  =Z_{\varphi}^{-1/2}\left(  K_{R}+\tau_{R}^{2}
/2v_{R}\right)=Z_{\varphi}^{-1/2}\overline{K}_{R},\label{Kbarr}\\
\overline{\tau}^{2}  &  =(\tau-\tau_{s})^{2}+2v(K-K_{s})\nonumber\\
& =Z_{\varphi}^{-2}Z_{v}\left(\tau_{R}^{2}+2v_{R}
K_{R}\right) =Z_{\varphi}^{-2}Z_{\tau}\overline{\tau}_{R}^{2}, \label{taubarr}%
\end{align}
One notices that both $\overline{K}$ and $\overline{\tau}$ are renormalized simply in
contrast with the inhomogeneity of the renormalization of $K$ itself. Note
that $\overline{\tau}^{2}$ is not renormalized simply as $\tau^{2}$. This underlies the difference between them.

The particular renormalization of $K$ in Eq.~(\ref{renormalization factors}) is to subtract divergent tadpole
contributions~\cite{Breuer,Vasilev}. Because
\begin{equation}
\Gamma^{(10)}(\{\mathbf{0},0\},\tau,u)=\lambda K,\quad\Gamma_{R}%
^{(10)}(\{\mathbf{0},0\},\tau_{R},u_{R})=\lambda_{R}K_{R}, \label{gamma10}%
\end{equation}
along with (\ref{renormalization factors}), they then leads to
\begin{align}
\Gamma_{R}^{(10)}(\{\mathbf{0},0\},\tau_{R},u_{R})=  &  Z_{{\widetilde{\varphi}}}^{1/2}%
\Gamma^{(10)}(\{\mathbf{0},0\},\tau,u)\nonumber\\
&  -Z_{0}\lambda_{R}\tau_{R}^{2}/2v_{R}. \label{factor 4}%
\end{align}
Indeed, after the mass renormalization, to one loop order, $\Gamma^{(10)}=\lambda
N_{d}v(\tau-\tau_s)^{2-\epsilon/2}/2\epsilon$, which is just canceled by the
subtraction in Eq.~(\ref{factor 4}) to the same order.

\subsection{Renormalization-group equations and their solutions for the general theory}
The scaling and universality behavior can be derived from an RG equation~\cite{Wilson,Ma,Justin,amitb,Kleinert,Vasilev}. In this section, we shall study the scaling behavior of the general theory~(\ref{action2}) to collect necessary ingredients for the reduced massless and purely massive theories in the next section. It will be seen that the solution to the RG equation automatically combines $K$ with $\tau$ according to Eq.~(\ref{Kbar}).

We first consider $\Gamma^{(1N)}$. At the instability point, $m\equiv\langle\varphi\rangle=0$. Exploiting the independency of the bare vertex on the momentum scale $\mu$, one finds
\begin{widetext}
\begin{equation}
  \left\{  \mu\frac{\partial}{\partial\mu}+\beta\frac{\partial}{\partial
u_{R}}+\gamma_{\lambda}\lambda_{R}\frac{\partial}{\partial\lambda_{R}}
+\overline{\gamma}_{\tau}\tau_{R}\frac{\partial}{\partial\tau_{R}}- \frac{1}{2}\widetilde{\gamma}-\frac{1}{2}N\gamma\right\} \Gamma^{(1N)}_{R}
  =\delta_{N0}\gamma_{\tau}\tau_{R}^{2}/2v_{R},\label{RG1}
\end{equation}
from Eqs.~(\ref{gnn}) and (\ref{factor 4}), where the Wilson functions are defined as derivatives at constant bare parameters as
\begin{align}
\gamma(u_{R}) & =\mu\frac{\partial\ln Z_{\varphi}}{\partial\mu},\qquad
& \overline{\gamma}_{\tau}(u_{R})   =\mu\frac{\partial\ln\tau_{R}}{\partial\mu
}=\gamma-\gamma_{\tau},\text{~~}\qquad
\gamma_{\lambda}(u_{R}) & =\mu\frac{\partial\ln\lambda_{R}}{\partial\mu} =\frac{1}{2}\widetilde{\gamma}-\frac{1}{2}\gamma, \nonumber\\
\widetilde{\gamma}(u_{R})  &  =\mu\frac{\partial\ln Z_{\widetilde{\varphi}}}{\partial\mu},\qquad
& \gamma_{\tau}(u_{R})  =\mu\frac{\partial\ln Z_{\tau}}{\partial\mu}=\mu\frac{\partial\ln Z_{v}}{\partial\mu},\qquad
\beta(u_{R})  & =\mu\frac{\partial u_{R}}{\partial\mu}=-u_{R}\left(
\frac{1}{2}\epsilon-\frac{3}{2}\gamma+\gamma_{\tau}\right)\label{Wilson}
\end{align}
with the help of Eqs.~(\ref{renormalization factors}) and (\ref{field 6}). The inhomogeneous term in Eq.~(\ref{RG1}) comes from the subtraction in Eq.~(\ref{factor 4}).

However, we are interested in the behavior near the instability point. For $r\neq r_c$, $m\neq 0$. The usual method is then to expand $\Gamma^{(10)}(m)$ at $m=0$. So, we have
\begin{equation}
\lambda_{R}K_{R}(\omega,\lambda_{R},\tau_{R},m_{R},u_{R},\mu)  =\Gamma_{R}%
^{(10)}(\omega,\lambda_{R},\tau_{R},m_{R},u_{R},\mu)=\sum_{N=1}^{\infty}
\frac{1}{N!}\Gamma_{R}^{(1N)}(\omega,\lambda_{R},\tau_{R},0,u_{R},\mu)m_{R}^{N},\label{RG field}\\
\end{equation}
where
\begin{equation}
m=Z_{\varphi}^{1/2}m_{R},\label{mzmr}
\end{equation}
similar to $\varphi$. Using Eqs.~(\ref{RG1}) and (\ref{mzmr}), we then obtain an inhomogeneous RG equation for $K_R$ as
\begin{equation}
\left\{  \mu\frac{\partial}{\partial\mu}+\beta\frac{\partial}{\partial u_{R}
}+\gamma_{\lambda}\lambda_{R}\frac{\partial}{\partial\lambda_{R}}+\overline{\gamma
}_{\tau}\tau_{R}\frac{\partial}{\partial\tau_{R}}-\frac{1}{2}\gamma
\left(1+m_{R}\frac{\partial}{\partial m_{R}}\right)\right\}  K_{R}(\omega,\lambda_{R},\tau_{R},m_{R},u_{R},\mu)=\gamma_{\tau}\tau_{R}^{2}
/2v_{R}.\label{RG2}
\end{equation}
This equation can also be obtained directly from Eq.~(\ref{renormalization factors}) if one assumes $K_R$ is a function of the other variables including $m_R$ that satisfies Eq.~(\ref{mzmr}).

Equation~(\ref{RG2}) is inhomogeneous. However, in terms of $\overline{K}_R$ similar to the definition in Eq.~(\ref{Kbar}) and used in Eq.~(\ref{Kbarr}), it can be rewritten in a homogeneous form as
\begin{equation}
\left\{  \mu\frac{\partial}{\partial\mu}+\beta\frac{\partial}{\partial u_{R}
}+\gamma_{\lambda}\lambda_{R}\frac{\partial}{\partial\lambda_{R}}+\overline{\gamma
}_{\tau}\tau_{R}\frac{\partial}{\partial\tau_{R}}-\frac{1}{2}\gamma
\left(1+m_{R}\frac{\partial}{\partial m_{R}}\right)\right\}  \overline{K}_{R}(\omega,\lambda_{R},\tau_{R},m_{R},u_{R},\mu)=0\label{RGKbar}
\end{equation}
by noting that
\begin{equation}
\left\{  \mu\frac{\partial}{\partial\mu}+\beta\frac{\partial}{\partial u_{R}
}+\gamma_{\lambda}\lambda_{R}\frac{\partial}{\partial\lambda_{R}}+\overline{\gamma
}_{\tau}\tau_{R}\frac{\partial}{\partial\tau_{R}}-\frac{1}{2}\gamma
\left(1+m_{R}\frac{\partial}{\partial m_{R}}\right)\right\} \frac{\tau_{R}^{2}}{2v_{R}}=-\gamma_{\tau}\frac{\tau_{R}^{2}}{2v_{R}},\label{RGtau2/2v}
\end{equation}
using Eqs.~(\ref{renormalization factors}) and (\ref{Wilson}) and combining Eqs.~(\ref{RG2}) and (\ref{RGtau2/2v}).

Solving Eq.~(\ref{RGKbar}) is then standard.
At the fixed point at which $\beta(u_{R}^{\ast})=0$, $\gamma$, $\gamma_{\lambda}$, and $\overline{\gamma
}_{\tau}$ become constants marked by stars, the solution is
\begin{equation}
\overline{K}_{R}(\omega,\lambda_{R},\tau_{R},m_{R},u_{R},\mu)
 =\kappa^{\frac{d+2-\gamma^*}{2}}\overline{K}_{R}\left(\omega/\lambda_{R}\kappa^{2+\gamma_{\lambda}^*},\tau
_{R}\kappa^{-2+\overline{\gamma}_{\tau}^*},m_{R}\kappa^{-\frac{d-2+\gamma^*}{2}},u_{R},\mu\right).\label{solve3}
\end{equation}
\end{widetext}
Employing the usual definitions of instability exponents,
\begin{align}
\eta &  =\gamma^{\ast},\qquad\qquad\beta/\nu=(d-2+\eta)/2,\nonumber\\
z  &  =\text{\ }2+\gamma_{\lambda}^{\ast},\qquad1/\nu=2-\overline{\gamma}_{\tau}^*,\nonumber\\
\beta\delta/\nu &  =\left(  d+2-\eta\right)  /2,\label{exponents}
\end{align}
we can write Eq.~(\ref{solve3}) in a familiar form,
\begin{equation}
\overline{K}_{R}(\lambda_{R},\tau_{R},m_{R})
  =\kappa^{\beta\delta/\nu}{{\overline{K}}}_{R}(\lambda_{R}t\kappa^{z},\tau
_{R}\kappa^{-1/\nu},m_{R}\kappa^{-\beta/\nu}),\label{form1}
\end{equation}
where we have used identical symbols for variables in both the time and the frequency domain. Near an instability point, we can solve $m_R$ from Eq.~(\ref{form1}) and obtain
\begin{equation}
m_{R}(\lambda_{R},\tau_{R},\overline{K}_{R})
  =\kappa^{\beta/\nu}m_{R}(\lambda_{R}t\kappa^{z},\tau
_{R}\kappa^{-1/\nu},{{\overline{K}}}_{R}\kappa^{-\beta\delta/\nu}).\label{form2}
\end{equation}

One sees that the scaling form~(\ref{form2}) is similar to Eq.~(\ref{mrho}) and $K_R$ and $\tau_R^2/2v_R$ are automatically combined together as a single variable in the solution. However, Eq.~(\ref{form2}) contains $\tau_R$ as an independent variable, which is incorrect. The RG theory cannot tell us why this variable must be absent. If we omit it, we have a correct scaling form that verifies Eq.~(\ref{mrho}).

From Eq.~(\ref{Wilson}), one finds that at the fixed point~\cite{Breuer},
\begin{equation}
\beta(u_{R}^{\ast})=-u_{R}^{\ast}\left(  \frac{\epsilon}{2}-\frac{3}{2}\gamma^{\ast
}+\gamma_{\tau}^{\ast}\right)  =0. \label{beta}%
\end{equation}
This gives rise to an exact relation,
\begin{equation}
\gamma_{\tau}^{\ast}=\frac{3}{2}\gamma^{\ast}-\frac{\epsilon}{2},
\label{gamma}%
\end{equation}
which reflects the relationship of the renormalization factors. As a result,
we have exactly
\begin{align}
1/\nu  &  =2-(\gamma^{\ast}-\gamma_{\tau}^{\ast})=2-(\epsilon
-\eta)/2,\nonumber\\
\beta &  =\nu(d-2+\eta)/2=1\label{standard}
\end{align}
from Eqs.~(\ref{Wilson}) and (\ref{exponents}).

Using Eq.~(\ref{zz}), we obtain
\begin{equation}
\beta(u_{R}^{\ast})=-\epsilon u_{R}^{\ast}/2-3u_{R}^{\ast3}/4=0.
\label{fixed point}%
\end{equation}
for the fixed points to one-loop order. One solution is the trivial Gaussian fixed point $u_{R}^{\ast}=0$. The other is $u_{R}^{\ast2}=-2\epsilon/3$, which is imaginary. However, it is infrared stable~\cite{zhongchen,Zhong2} for $\epsilon>0$ similar to the fixed point for critical phenomena~\cite{Justin,amitb}. From Eqs.~(\ref{zz}), (\ref{Wilson}), (\ref{exponents}), and (\ref{standard}), one obtains~\cite{zhongchen,Zhong2}
\begin{align}
\eta & =-\epsilon/9,\qquad ~1/\nu = 2-5\epsilon/9,\nonumber\\
z & =2-\epsilon/18, \qquad\delta  =2-7\epsilon/18,\label{expe}
\end{align}
to one loop. All exponents are real. For $\epsilon=0$, all the exponents recover their respective mean-field values correctly.

\subsection{Renormalization-group equations and their solutions for the reduced theories}
As we have seen from last section, the general theory contains one redundant parameter which has to be removed by hand~\cite{Vasilev}. In this section, we study the reduced theories that directly give rise to correct results.

Because of the relation~(\ref{tauc1}), we can either shift the mass to the field and obtain the massless theory or vice versa and arrive at the purely massive theory similar to the mean-field approximation.

We start with the massless theory, Eq.~(\ref{Langevin 2}) with the Gaussian noise~(\ref{noise}). Its dynamic functional is similar to Eq.~(\ref{action2}) in the absence of $\tau$ and with $\overline{K}$ replacing $K$. Its RG equation is then simply
\begin{align}
\left\{   \mu\frac{\partial}{\partial\mu}  +\beta\frac{\partial}{\partial
u_{R}}\right.&\left. +\gamma_{\lambda}\lambda_{R}\frac{\partial}{\partial\lambda_{R}}\right.\nonumber\\
&\left.-\frac{1}{2}\gamma\left(1+m_R\frac{\partial}{\partial m_{R}}\right)\right\}\overline{K}_{R}
=0 \label{RG5}
\end{align}
similar to Eq.~(\ref{RGKbar}). Its solution at the infrared-stable fixed point is
\begin{equation}
{{\overline{K}}}_{R}(\lambda_{R},m_{R},u_{R}^{\ast})=\kappa^{\beta\delta/\nu}%
{{\overline{K}}}_{R}(\lambda_{R}t\kappa^{z},m_{R}\kappa^{-\beta/\nu},u_{R}^{\ast}),
\label{form3}%
\end{equation}
or, in terms of $m_R$,
\begin{equation}
m_{R}(\lambda_{R},\overline{K}_{R})
  =\kappa^{\beta/\nu}m_{R}(\lambda_{R}t\kappa^{z},{{\overline{K}}}_{R}\kappa^{-\beta\delta/\nu}),\label{mkbar}
\end{equation}
which is the scaling form~(\ref{mrho}) with a momentum instead of a length rescaling factor.

Now we turn to the purely massive theory, Eq.~(\ref{Langevin 3}) with the Gaussian noise~(\ref{noise}). Its dynamic functional is again similar to Eq.~(\ref{action2}) now in the absence of $K$ and with $\overline{\tau}$ replacing $\tau$. Its RG equations can be obtained from $m_{R}=G_{R}^{(10)}
(\{{\mathbf{0}},0\};\lambda_{R},\tau_{R},u_{R},\mu)$. It is
\begin{equation}
\left\{  \mu\frac{\partial}{\partial\mu}+\beta\frac{\partial}{\partial
u_{R}}+\gamma_{\lambda}\lambda_{R}\frac{\partial}{\partial\lambda_{R}} +\gamma_{\overline{\tau}}\overline{\tau}_{R}\frac{\partial}
{\partial\overline{\tau}_{R}}+\frac{1}{2}\gamma\right\}  m_{R}
 =0,\label{RG6}\\
\end{equation}
with an additional Wilson function
\begin{equation}
\gamma_{\overline{\tau}}(u_{R})=\mu\frac{\partial\ln\overline{\tau}_{R}}{\partial\mu
}=\gamma-\frac{1}{2}\gamma_{\tau}\label{taubarr1}
\end{equation}
from Eqs.~(\ref{taubarr}) and (\ref{Wilson}). Equation~(\ref{RG6}) can also be obtained by assuming $m_R$ is a function of the other variables. The fixed-point solution is then
\begin{equation}
m_{R}(\lambda_{R},\overline{\tau}_{R},u_{R}^{\ast})=\kappa^{\beta/\nu
}m_{R}(\lambda_{R}t\kappa^{z},\overline{\tau}_{R}\kappa^{-1/\overline{\nu}}
,u_{R}^{\ast}),\label{form4}%
\end{equation}
with
\begin{equation}
1/\overline{\nu}=2-\gamma^{\ast}+\gamma_{\tau}^{\ast}/2\label{theta}
\end{equation}
from the fixed-point solution of Eq.~(\ref{taubarr1}). Equation~(\ref{form4}) is just another scaling form of (\ref{mrt}) and thus confirms the latter. Using the exact relation~(\ref{gamma}) and the definition~(\ref{exponents}), we find
\begin{equation}
\overline{\nu}=4/\left(  d+2-\eta\right)=2\nu/\beta\delta.\label{nub}
\end{equation}
This is just the relation that we obtained from the scaling theories and promised to prove.

Relation~(\ref{nub}) is, as mentioned, surprising as $\overline{\tau}$ appears proportional to $\tau$ from its definition~(\ref{taubar}). On the other hand, however, it may be expected as $\overline{\tau}^2$ is proportional to $\overline{K}$ again from their definitions and thus shares with the latter identical anomalous dimensions. Indeed, from Eqs.~(\ref{form4}) and (\ref{nub}), the exponent for $\overline{\tau}^2$ is just $2/\overline{\nu}$ and is thus $\beta\delta/\nu$, the exponent for $\overline{K}$, as seen from Eq.~(\ref{mkbar}). The proportional coefficient, $2v$, between $\overline{\tau}$ and $\overline{K}$, as seen from Eqs.~(\ref{Kbar}) and (\ref{taubar}), just compensates their difference in their na\"{i}ve dimensions.

Using Eqs.~(\ref{standard}) and (\ref{expe}), we find
\begin{equation}
\overline{\nu}=\frac{1}{2}\left(1+\frac{17}{36}\epsilon\right)\label{nue}
\end{equation}
to one-loop order. The mean-field result is thus $1/2$, equal to $\nu$ as mentioned.

For the purely thermal models including the generalized one, one just needs to replace $\overline{\tau}$ with $\tau$ itself. The fixed-point solution is simply
\begin{equation}
m_{R}(\lambda_{R},\tau_{R},u_{R}^{\ast})=\kappa^{\beta/\nu
}m_{R}(\lambda_{R}t\kappa^{z},\tau_{R}\kappa^{-1/\nu}
,u_{R}^{\ast}),\label{form5}%
\end{equation}
which is just Eq.~(\ref{form2}) of the general theory without $\overline{K}$. However, as has been pointed out in Sec.~\ref{fluctuations}, the purely thermal model with $\sigma=2$ itself does not describe correctly the temperature-driven FOPTs. For the generalized model, the renormalization factors are different from the $\varphi^3$ ones, though the symbols for the exponents are identical.

\subsection{Exponents of thermal hysteresis}

The RG theories have confirmed the scaling forms developed in Sec.~\ref{scaling}. They also yield the $\epsilon$-expansions for the hysteresis exponents. The relation~(\ref{nub}) has also confirmed the identity of the two sets of the exponents obtained from the two reduced theories. So, we shall only consider the hysteresis exponents of the massless theory, Eqs.~(\ref{nTm}), (\ref{nT}), and (\ref{nTtk1}), and the purely thermal theory, Eq.~(\ref{nTt3}).

Using Eqs.~(\ref{exponents}) and (\ref{standard}), we can write the hysteresis exponents for the three thermal classes as
\begin{eqnarray}
\Upsilon  &  =&\frac{\beta\delta}{\beta\delta+\nu z}=\frac{d+2-\eta}{d+2-\eta+2z},\label{N1}\\
\Upsilon_{m}  &  =&\frac{\beta}{\beta\delta+\nu z}=\frac{d-2+\eta}{d+2-\eta+2z},\label{N2}\\
\dot{\Upsilon} &=& \frac{\beta\delta}{\beta\delta+2\nu z}=\frac{d+2-\eta}{d+2-\eta+4z},\label{N3}\\
\dot{\Upsilon}_{m} &  =& \frac{2\beta}{\beta\delta+2\nu z}=\frac{2(d-2+\eta)}{d+2-\eta+4z},\label{N4}\\
\ddot{\Upsilon}&=&\ddot{\Upsilon}_{m}=\frac{d-2+\eta}{d-2+\eta+2z},\label{N5}
\end{eqnarray}
To one-loop order, they becomes
\begin{align}
\Upsilon  &  =\frac{2}{3}\left(1-\frac{1}{36}\epsilon\right),\quad &\dot{\Upsilon} = \frac{1}{2}\left(1-\frac{1}{24}\epsilon\right),\nonumber\\
\Upsilon_{m}  &  =\frac{1}{3}\left(1-\frac{7}{36}\epsilon\right),\quad
&\dot{\Upsilon}_{m}   = \frac{1}{2}\left(1-\frac{5}{24}\epsilon\right)\label{N4e}
\end{align}
from Eq.~(\ref{expe}). We have not listed $\ddot{\Upsilon}$ and $\ddot{\Upsilon}_{m}$ as they do not describe the temperature-driven transitions for $\sigma=2$. Nevertheless, to one-loop order, they are $\ddot{\Upsilon} = \ddot{\Upsilon}_{m}=1/2-\epsilon/18$, which are indeed different from those of Thermal Class II beyond mean-field level. The field-like hysteresis exponents are in fact identical to the field-driven FOPTs as mentioned and have been given in Ref.~\cite{Zhong2}. The thermal-like exponents are the results of the present study.

\begin{table*}[ht]
\caption{\label{exponent2}Instability exponents and thermal hysteresis exponents\footnote{quoted errors reflect the spread in different resummations except the last two rows}}
\begin{ruledtabular}
\begin{tabular}{lllllllll}
d &  & $6$ & 5 & 4 & 3 & 2 & 1 & 0\\
\hline
$\eta$&\cite{Zhong2} &0&$-0.147\pm0.002$&$-0.329^{+0.012}_{-0.013}$&$-0.527^{+0.029}_{-0.033}$ &$-0.747^{+0.064}_{-0.050}$ &$-1$ \cite{Fisher}&$-1.224$\\
$z$&\cite{Zhong2}&2&\quad\!$1.941\pm0.003$&\quad\!$1.880\pm0.006$&\quad\!$1.817\pm0.008$&\quad\!$1.753\pm0.010$&\quad\!$1.677$ &\quad\!$1.612$\\
$\Upsilon $ & Thermal Class I~\cite{Zhong2} & 2/3 & \quad\!$0.6480\pm0.0004$ &
\quad\!$0.627\pm0.001$ &\quad\!$0.603\pm0.003$ &\quad\!$0.575\pm0.004$ & \quad\!0.544 & \quad\!1/2\\
$\Upsilon_{m}$ & Thermal Class I~\cite{Zhong2} & 1/3 & \quad\!$0.259\pm0.0003$ &
\quad\!$0.166\pm0.002$ & \quad\!$0.0516^{+0.0031}_{-0.0035}$ &$-0.0905^{+0.0047}%
_{-0.0062}$ & $-0.272$ & $-1/2$\\
$\dot{\Upsilon}$ & Thermal Class II & 1/2 &\quad\!$0.4793\pm0.0005$ &\quad\!$0.457\pm0.002$ &\quad\!$0.432\pm0.003$ & \quad\!$0.404\pm0.004 $&\quad\!0.374 &\quad\!1/3\\
$\dot{\Upsilon}_{m}$ & Thermal Class II & 1/2 &\quad\!$0.3827\pm0.0005$ &\quad\!$0.241^{+0.002}_{-0.003}$ &
\quad\!$0.0739^{+0.005}_{-0.006}$ &$-0.127^{+0.007}_{-0.009}$ & $-0.374$ & $-2/3$\\
$\ddot{\Upsilon}=\ddot{\Upsilon}_{m}$ & Thermal Class III & 1/2 &\quad\!---&\quad\!--- &\quad\!--- &\quad\!--- &\quad\!--- &\quad\!---\\
$\ddot{\Upsilon}$ & $\sigma=3$~\cite{zhonge} & 1/2 &\quad\!1/2 &\quad\!1/2 &\quad\!$0.4380(15)$ &\quad\!$0.3158(1) $ & & \\
$\ddot{\Upsilon}_{m}$ & $\sigma=3$~\cite{zhonge} & 1/2 &\quad\!1/2 &\quad\!1/2 &\quad\!$0.1430(49)$ &\quad\!$0.03948(2) $ & & \\
\end{tabular}
\end{ruledtabular}
\end{table*}
The estimates of the hysteresis exponents can be further improved. By utilizing the three- and two-loop results of the static~\cite{Alcantara} and dynamic~\cite{Breuer} exponents for the Yang--Lee edge singularity and some exact results in low dimensions~\cite{Fisher}, Pad\'{e} resummations have been performed and the best estimates for $\eta$ and $z$ up to date have been given~\cite{Zhong2}. From these results, similar estimates for the hysteresis exponents in Eqs.~(\ref{N1}) to (\ref{N5}) can also be computed. These are all listed in Table~\ref{exponent2}.
Included in the last two rows are the results of the general purely thermal class for $\sigma=3$, the $\phi^6$ model studied in Sec.~\ref{gptc}. Because of the same universality class, these two exponents are extracted directly from Ref.~\cite{zhonge}, where they are the hysteresis exponents for nonequilibrium \emph{critical} phenomena of the Ising universality class.

The mean-field hysteresis exponents listed in Table~\ref{exponent2} have been confirmed numerically. The hysteresis exponents listed explicitly in $d=0$ are exact because in this dimension, $z=(2-\eta)/2$~\cite{Breuer}. The hysteresis exponents for the purely thermal class with $\sigma=3$ in $d=2$ have been verified numerically~\cite{zhonge} and so have those for the field-like class in $d=7$ to 4~\cite{Yu}. The remaining hysteresis exponents have yet to be tested.

Experimentally, the mean-field exponent $\Upsilon$ has been confirmed in liquid crystals~\cite{Yildiz}. It has also been estimated to ranging from $0.26$ to $0.49$ in a couple of alloys and compounds~\cite{Lin,Zhang,liu,fung,Kuang}, not far away from the theoretically value in $d=3$. However, the ranges of the sweep rate $R$ employed in these experiments are small and thus further experiments are desirable.

\section{\label{summary}SUMMARY}

We have studied the scaling and universal behavior of temperature-driven first-order phase transitions (FOPTs). We have shown that these transitions exhibit rich phenomena though they are controlled by a single complex-conjugate pair of the imaginary fixed points of the $\phi^3$ theory.

The expansion near the spinodal or instability point of an FOPT results in a leading $\phi^3$ theory. Its shift symmetry leads to two equivalent reduced models, the massless model~(\ref{Langevin 2}) and the purely massive model~(\ref{Langevin 3}). Although the latter is real only in the absence of the external field and under cooling, in which case it becomes the purely thermal model~(\ref{ptm}) or its generalizations~(\ref{Langevin 4}), it falls into the same universality class to the massless model and verifies the unique properties of the $\phi^3$ theory. Scaling theories have also been proposed. The resultant scaling forms give rise to several universality classes with their own hysteresis exponents. These include the field-like Thermal Class I, the partly thermal class Thermal Class II, and the purely thermal class Thermal Class III. The first two classes are opposite limits of the scaling forms and may cross over to each other depending on the temperature sweep rate $R$. They are both described by the massless model and the purely massive model. The last class is characterized by the purely thermal models and contains different universality classes depending on the symmetry of the order parameters. An example is the $\phi^6$ model whose cooling transition in the absence of an applied external field falls into the same universality class to the nonequilibrium critical phenomena of a usual $\phi^4$ model. If odd-symmetry terms are allowed in the free energies, Thermal Class III emerges only in the mean-field limit and is identical with Thermal Class II. It changes to the other two classes when fluctuations are considered. Numerical and analytical results in the mean-field level agree well with the scaling analysis. The renormalization-group theories both confirm the scaling theory and the relation between the massless model and the purely massive model and provide methods to calculate the hysteresis exponents of various universality classes. Using the extant three- and two-loop results for the static and dynamic exponents for the Yang-Lee edge singularity, which falls into the same universality class to the $\phi^3$ theory, we have estimated the thermal hysteresis exponents of the various classes to the same precisions. A few exponents have already been verified both numerically and experimentally and further comparisons are desirable.

\section*{acknowledgments}
We thank Shuai Yin and Baoquan Feng for their helpful discussions. This work was supported by the National Natural Science foundation of PRC (Grants No. 10625420 and 11575297) and FRFCUC.

\end{document}